\documentclass[prc,nofootinbib,showpacs,twocolumn]{revtex4}
\usepackage{graphicx}
\usepackage[usenames,dvipsnames]{color}
\usepackage{amsmath,amssymb,bbold}
\usepackage{hyperref}
\begin{document}
\title{Setting Initial Conditions for Inflation with
Reaction-Diffusion Equation}
\author{Partha Bagchi $^{1,3}$ \footnote {email: p.bagchi@vecc.gov.in}}
\author{Arpan Das$^{2,3}$ \footnote {email: arpan@iopb.res.in}}
\author{Shreyansh S. Dave$^{2,3}$ \footnote {email: shreyansh@iopb.res.in}}
\author{Srikumar Sengupta $^{2,3}$ \footnote {email: srikumar@iopb.res.in}}
\author{Ajit M. Srivastava$^{2,3}$ \footnote{email: ajit@iopb.res.in}}
\affiliation{$^1$ Variable Energy Cyclotron Centre, 1/AF, Bidhannagar,
Kolkata - 700 064, India\\
$^2$ Institute of Physics, Bhubaneswar 751005, India\\
$^3$ Homi Bhabha National Institute, Training School Complex,
Anushakti Nagar, Mumbai 400085, India}
\begin{abstract}
 We discuss the  issue of setting appropriate initial conditions for 
inflation. Specifically, we consider natural inflation model and
discuss the fine tuning required for setting almost homogeneous
initial conditions over a region of order several times the Hubble 
size which is orders of magnitude larger than any relevant correlation
length for field fluctuations. We then propose to use the special
propagating front solutions of reaction-diffusion equations for
localized field domains of smaller sizes. Due to very small
velocities of these propagating fronts we find that the inflaton
field in such a {\it field domain} changes very slowly, contrary to
naive expectation of rapid roll down to the true vacuum. Continued
expansion leads to the energy density in the Hubble region being
dominated by the vacuum energy, thereby beginning the inflationary
phase. Our results show that inflation can occur even with a single
localized field domain of size smaller than the Hubble size.
We discuss possible extensions of our results for different 
inflationary models, as well as various limitations of our analysis 
(e.g. neglecting self gravity of the localized field domain). 
\end{abstract}
\maketitle

\section{INTRODUCTION}

The hot big-bang model of the Universe is described by 
Friedmann-Robertson-Walker (FRW) metric \cite{weinberg6}, which 
is based on large-scale homogeneity and isotropy of the observed universe. It 
provides reliable and tested description of the history of the Universe from 
about 1 sec after the big-bang till today.  Despite the self-consistency and 
remarkable success of the early hot big-bang model, a number of unanswered 
questions remained regarding the initial state of the Universe. These are 
known as flatness problem, horizon problem and monopole problem 
\cite{riotto6}. To resolve these problems inflationary universe model was 
proposed. The main ingredient of the inflation is the accelerated expansion 
of the Universe at a very early stage. If the duration of the inflation 
$\Delta t\geq 60 H^{-1}$, where $H$ is the Hubble parameter 
during inflation, then one can explain the observed homogeneity and isotropy 
of the Universe, the absence of the magnetic monopoles and the spatial 
flatness. Although it was known that inflation can resolve the shortcomings
of the hot big-bang cosmology, the first particle physics motivated model of 
inflation was suggested by Guth \cite{guth6}, 
which is known as the old inflationary model. In this model, the Universe 
is assumed to be initially in thermal equilibrium and undergoes a strong 
first order phase transition (typically at the GUT scale). 
Inflation occurs when the scalar field (inflaton) gets trapped in the
metastable vacuum. In this model inflation ends by quantum tunneling of the 
field into the true  vacuum via bubble nucleation. In 
his own paper, Guth pointed out the graceful exit problem of this model. In 
old inflation model, the Universe expands at an exponential rate while the 
bubbles nucleate at a constant rate. Until and unless the nucleation 
rate is high enough, the bubbles will not collide with each other, which is 
necessary for the end of the inflation and subsequent reheating. However, if 
the nucleation rate is large then phase transition will be completed 
quickly and it will not give enough inflation. Thus old inflationary model 
suffers graceful exit problem. This problem was addressed in the new inflation 
model \cite{newinfl61,newinfl62,newinfl63}, where the shape of the 
finite temperature effective potential is such that either there is no 
potential barrier, so the phase transition is second order, or there
is a very tiny potential barrier. In this model, the potential 
must have a very flat portion in between the origin and the true minimum. 
The field starts very close to $\phi = 0$ and slowly rolls to the true 
minimum. During the slow roll, the energy density is dominated by the 
potential energy which gives rise to inflation. In new inflation when the 
classical field reaches the end of the flat part of the potential, it rapidly
rolls down to the true vacuum and starts to oscillate. This is the stage of
reheating of the Universe at the end of inflation. Other inflationary models 
are also proposed like chaotic inflation\cite{chaotic6}, stochastic 
inflation \cite{stochastic6}, inflation with pseudo  Goldstone 
boson \cite{natural61}, inflation with modified gravity \cite{fr6}, warm
inflation \cite{warm6}etc.

 Inflationary models can be broadly classified depending upon initial 
conditions, behavior of the scale factor and the end of the inflation.
(Warm inflation is a separate category where dissipative dynamics of
the inflaton field leads to continued particle production during the
entire inflationary phase. We will comment on that later.) For example, 
the old and new inflationary models were based on the assumption that the 
Universe was in a state of thermal equilibrium at very high temperature, on 
the other hand, chaotic inflation models can be applied for generic 
initial conditions including no thermal equilibrium for the 
pre-inflationary stage. Again time dependence of 
the scale factor during inflation can be different in different models e.g. 
exponential inflation, power-law inflation etc. Inflationary models can also 
be classified according to the change in the scalar field during inflation,
e.g. the large field and small field models where the change in the 
scalar field is of the order of $m_{pl}$ (Planck mass), or smaller
respectively (see, ref.\cite{linde} for different definitions used 
in the literature for the term large field inflation).

  In the original old inflation model, although the model does not work due 
to graceful exit problem, $\phi =0$ is naturally set because of the presence 
of the metastable vacuum at $\phi = 0$. For new inflation it was hard
to justify initial value of field being close to zero when the transition
is second order (which requires field always remaining at the minimum of 
the potential) \cite{new1}. For this, versions 
of new inflation invoked a metastable 
vacuum with tiny potential barrier, along with a very flat top for the 
potential \cite{newinfl61,newinfl62,bubl}. With that, the initial value of 
$\phi$ could be set precisely (e.g. equal to zero) due to the false vacuum 
at that value of $\phi$. The field tunnels through 
the barrier via nucleation of a bubble which undergoes inflation as the 
field inside the bubble rolls slowly over the flat top of the potential.
The inflation ends when the field reaches the end of flat part of the potential
and rolls down to the true vacuum rapidly.  Thus the entire observed universe 
is inside one very large bubble, with other such bubbles constituting
different parts of the Universe disconnected from our universe by 
regions which are constantly undergoing inflation. For other models like
chaotic inflation and  natural inflation, the initial value of the field 
is not set in this manner. Rather, it is supposed to explore 
entire allowed (relevant) range of field values. Inflation occurs wherever 
the field has the correct initial value. 

For definiteness we will discuss natural inflation model with field varying 
from 0 to the vacuum expectation value (vev). 
For inflation, (e.g. for natural inflation with the field
starting near $\phi = 0$) the requirement of appropriate field values 
over several Hubble volumes is very hard to justify. We will discuss this 
in some detail in the next section. Important point is that correct field 
value is required not just as an average, one cannot allow significant field 
values anywhere inside the region of interest (several Hubble volumes) as the 
field will rapidly roll down to the true vacuum in those regions.
The issue of initial conditions for inflaton field has been extensively 
discussed in the literature 
\cite{berera1,berera2,robert,robertrev,east26,intl}. In 
particular, it has been investigated that starting 
with inhomogeneous field  configurations, how a homogeneous component,
appropriately localized near the origin can arise. Recently, 
fluctuation-dissipation dynamics has been used in this context 
by Bastero-Gil et al. \cite{berera2}. This is of importance to us as
dissipative dynamics plays very crucial role in our model also. However, 
there are crucial differences between this work and our model , as
we will discuss below. In a  numerical simulation of full 
Einstein  field equations coupled to scalar inflaton field, it is shown 
by East et al. 
\cite{east26} that inflation can occur with highly inhomogeneous initial
conditions, as long as the average value of the field, as well as
the fluctuation of the field in a Hubble region 
remains within the slow roll region of the potential. Otherwise, as we 
discussed above, the field generically rolls down quickly to the minimum, 
being pulled down by the field values outside the slow roll regime.
For earlier works with highly inhomogeneous conditions, see 
ref. \cite{berera1,robert}.  For a recent review on the 
issue of initial conditions, see ref.\cite{robertrev} (see, also 
\cite{raghu}).

 We attempt to address this issue of initial conditions for inflation using 
specific features of the reaction-diffusion (RD) equations 
\cite{reacdiff6,hep,our16,our26,skthesis}
using a natural inflation model with field varying between 0 and vev. 
In our model, the field is close to zero only in a very 
small region inside the Hubble volume and 
varies smoothly to a large value (vev) over a region, which we will call as
the {\it field domain}. The size of this {\it field domain} is taken to be
smaller than the Hubble size at GUT scale. In principle, it may be
possible to take the size of the {\it field domain} to be of order of
the correlation size, but in the present paper we are only able to
provide results when it is much bigger than the correlation domain, though
still smaller than the Hubble region at GUT scale. Domain sizes we
take provide examples where inflation can be shown to occur. Due
to long time required for each simulation we have not been able to
optimize the domain sizes for different cases. Important thing
is that the field is not homogeneous at all in our model, and 
the variation of the field lies well outside the slow 
roll region, in fact all the way upto the vacuum value.  
One would have expected this field to rapidly roll down
to the true vacuum everywhere. This is where the special features of
reaction-diffusion equation become relevant. With appropriate boundary
conditions, with field interpolating between $\phi = 0$ at the top of the
potential and the minimum of the potential (the true vacuum), specific 
solutions of reaction diffusion equation exist which represent slowly
propagating fronts just like interfaces in a first order transition 
\cite{reacdiff6,our16,our26,skthesis}.
This happens even though the potential corresponds to a 
second order transition, or even a cross-over.
This holds up the field near $\phi = 0$ for sufficiently long
time such that vacuum energy starts dominating compared to
the kinetic energy (from the time derivative of the field), the gradient 
energy, as well as the radiation energy,
in the entire Hubble volume signaling beginning of inflation. Duration of
inflation will be governed by the exact profile of the field inside this
field domain, and its size. 

Though still we are unable to show inflation
beginning from a single correlation-size domain, it is important to
appreciate that we are able to obtain inflation using a reasonable
field profile over a region smaller than the GUT scale Hubble region.
In fact in some cases we are even able to get field domain to be smaller
than the Hubble size including energy density of the field variation
(which has large contributions from the gradient energy).  
Important point is that one can get inflation even when the field profile
inside the Hubble volume is highly inhomogeneous with field varying from
the top of the potential to the vev. All that is required is
that contributions of field kinetic energy as well as  gradient energy, 
along with background radiation
energy density etc., become sub-dominant compared to the potential energy 
contribution. Even a single small field domain can achieve that if it 
retains its shape during evolution. Normally one expects that any 
inhomogeneous field will quickly roll down to true vacuum. However, as we 
will see, special solutions of the RD equations provide field profiles which 
retain their shape for very long time, allowing inflation to occur.
We emphasize that we take natural inflation model here just as an 
example to illustrate the importance of RD equation solutions in 
providing initial conditions for inflation. 
Our discussion naturally applies to any  inflation model with such
boundary conditions, and  can also be straightforwardly extended 
to other models if the potential has appropriate extrema. The only
requirement for the applicability of RD analysis is that the potential 
should have one maximum and a minimum, the corresponding values of 
field providing appropriate boundary conditions for the RD front
profile. The non-trivial role
of RD equation solutions in delaying the roll down of the field from
the top of the potential clearly has very general applicability and
should be explored for a wide class of inflation models.

 Small field domains with the field profile as discussed above
 will be expected to arise naturally during a thermal phase transition.
 For example, in a second order transition (or after a rapid first order
 transition), due to thermal fluctuations, small regions will generically 
 be found where the field remains near the top of the potential while 
 the field settles to the true vacuum in the neighboring regions. This
 is especially true for a continuous transition in the critical regime
 where fluctuations of a correlation domain to the top of the potential 
 are unsuppressed with the correlation length becoming very large.
 (Though, with rapid expansion at the GUT scale, large correlation domains
 may not be realistic due to critical slowing down.)
 Once such a field profile is achieved, and the universe 
 cools down below the Ginzberg temperature, rest of the dynamics of the 
 field will be governed by RD equation as we discussed above. One may
 expect similar small domains to arise from quantum fluctuations also 
 for a non-equilibrium transition (say a quench), or possibly even 
 without any thermal history of the universe just as one expects
 in the conventional picture of inflation, except that one needs much
 smaller domains here. (This is important as achieving thermal
 equilibrium for the inflaton field during pre-inflation stages is far 
 from clear in view of various constraints on the couplings \cite{turner}. 
 We briefly comment on this issue further in Sect.VI.
 Effects of any pre-inflation equilibrium stage on CMBR has been discussed 
 in ref.\cite{raghu2}.)

  We mention here one of the important limitations of our analysis.
We have not taken into account the self gravity of the field domain
for its evolution. The field in this region is assumed to undergo
evolution by the FRW equations appropriate for the average energy density
in that Hubble region even though energy density is highly inhomogeneous. 
However, for the highly inhomogeneous case we are considering, we cannot 
say whether the evolution of the whole domain does not change 
significantly with a complete treatment of inhomogeneity.  The evolution 
of such a localized field domain has to be studied using full Einstein 
equations in that region like the analysis in \cite{east26}. 
In that sense our results should be taken in the spirit of 
providing a new possibility of beginning the inflationary phase which needs
to be verified by more detailed simulations. Our main purpose in this
work is to demonstrate the role of RD equation solutions in slowing
down the roll down of the field from the top of the potential even when
spatial derivatives of the field are significantly 
non-zero.  Another limitation
of our model is that in studying the evolution of the field profile, 
though dissipation plays an important role, fluctuation part has been
ignored. Fluctuations can clearly affect the propagation of the fronts,
and should be considered (as we noted in our earlier works 
(\cite{our16,our26,skthesis}). For example, in \cite{berera2}, the role of 
fluctuation-dissipation dynamics has been 
found to play crucial role in setting the homogeneous component of the 
field appropriately near the origin. However, there are important differences
in that work and our model as role of special solutions of field
equation (namely RD equations here) along with proper boundary conditions, 
play no role in the analysis of ref.\cite{berera2}.
Though we have considered high dissipation here, with special
propagating front solutions of RD equation having very small velocities,
even smaller dissipation may work in our model. Highly inhomogeneous
initial conditions have also been considered by Albrecht et al. 
\cite{robert} where role of dissipation has been emphasized (see, also
\cite{berera1}). However,
role of special boundary conditions and special solutions of
propagating front (which may allow for smaller dissipation) were not
considered there.

 The paper is organized in the following manner. In Section II we will 
revisit the issue of requirement of initial fine tuned field value
for natural inflation. We will argue that it appears virtually impossible to 
satisfy this requirement over a Hubble volume. In Section III we will 
introduce the Reaction-diffusion equation where we will also
briefly mention our earlier application of RD equation in the context of 
heavy ion collision experiment and formation of misaligned chiral field
domains, the so called {\it disoriented chiral condensate} (DCC) in pp 
collisions at LHC energies. The reason we present this discussion is
that the physics of 
DCC domain formation using RD equation in the context of chiral sigma model 
very closely resembles the issue of initial condition for inflation. In
fact in our model, DCC formation is exactly equivalent to an inflating
field domain in the expanding universe. In some ways, this picture of
DCC formation in heavy-ion collisions can be taken as possible 
experimental test of the basic picture underlying the model we are 
proposing for inflation based on RD equation solutions. Section IV discusses 
application of RD equation solutions to the case of inflation in the early 
universe. Section V presents numerical results and 
conclusions are presented in Section VI.

\section{INITIAL CONDITIONS FOR INFLATION}
 
As we mentioned above, we will discuss the case of a natural inflation 
model  with pseudo Nambu-Goldstone 
boson inflaton field \cite{natural61}. Nambu- Goldstone bosons naturally 
arise in particle physics models with a spontaneous breaking of some global
continuous symmetry. If there is explicit symmetry breaking in addition to 
spontaneous symmetry breaking then one gets Pseudo-Nambu-Goldstone Bosons 
(PNGB). In the case of QCD axion model with Pecci-Quinn symmetry breaking 
two very different scales naturally arises, these are Pecci-Quinn scale 
$f_{PQ}$ which can be as high as the GUT scale $\sim 10^{15}GeV$,
and QCD scale $\Lambda_{QCD}\sim 200 MeV$. In this case axion self-coupling is
$\lambda_a\sim (\Lambda_{QCD}/f_{PQ})^4 \sim 10^{-64}$. The same situation 
arises in low energy QCD theory with chiral sigma model, where pions are PNGB. 

For natural inflation, we take the potential  of the form 
$V(\phi)=\Lambda^4(1\pm cos(\phi/f))$. We take the plus sign and take
the inflation to occur when $\phi$ rolls from a value near $\phi = 0$ to 
the vev ($\phi = \pi f$). (Our analysis will equally well apply to other
choices of potential maximum. See, for example the analysis of DCC formation
with chiral sigma model in ref. \cite{our26}.) 
This potential contains two scales 
which naturally comes from particle physics model. One can show that if 
$f\sim m_{pl}$ and $\Lambda \sim m_{GUT}\sim 10^{15} GeV$, then the 
PNGB field $\phi$ can drive inflation. 
Note $\Lambda$ is the scale of the potential and $f$ is the vacuum expectation 
value of the field after the symmetry breaking, which also set the scale of 
the field $\phi$. The values of $f$ and $\Lambda$ are constrained 
by the requirements of the slow-rolling regime, sufficient inflation and 
observed magnitude of density contrast of CMBR. In the temperature range $T\le 
f$, the global symmetry is spontaneously broken and Nambu-Goldstone boson 
describes the angular degree of freedom of the symmetry broken potential. 
Initially, the value of NGB is distributed uniformly between 0 and $2\pi f$ in 
different causally connected regions, or rather, different correlation
volumes. At temperature range $T\le \Lambda$ 
additional explicit symmetry breaking comes into the picture because of 
which  the original {\it Mexican Hat} potential becomes 
tilted. Now the PNGB field $\phi$  starts rolling down to 
the unique true vacuum. If one starts from any 
arbitrary value of $\phi$ then sufficient inflation is not guaranteed. 
Although one can get sufficient inflation if the initial field value is 
localized suitably near $\phi =0$.

In order to achieve sufficient inflation, the PNGB field has to start
close to $\phi = 0$ and one assumes that somewhere in the entire universe
such a condition was achieved initially in a region which is at least of Hubble
size (more like several times the Hubble size). Let us examine how reasonable
this requirement is. For definiteness, let us assume the Universe to be  
in equilibrium and radiation dominated before inflation. 
Inflaton field fluctuations will be
determined by the correlation length. Let us consider a GUT scale inflation 
at, say, 10$^{16}$ GeV scale. If one assumes that the Universe was radiation 
dominated between the Planck epoch and the GUT scale, then at the GUT scale 
$H^{-1}_{GUT}=\frac{T_{pl}^2}{T_{GUT}^2}H^{-1}_{pl}$ (though it may be hard
to argue that the inflaton field can be in equilibrium above the GUT scale 
\cite{turner}). We take the equilibrium
correlation length at the GUT temperature scale, $\zeta_{GUT}$ to be of the 
order of the inverse of the temperature, $\zeta_{GUT} = (T_{GUT})^{-1}$.
Ignoring factor of order 1, with $H_{pl} \simeq T_{pl}$, we find that
the Hubble volume at the GUT scale contains about $10^9$ uncorrelated 
domains ($10^{12}$ domains if GUT scale is taken to be $10^{15}$ GeV). 
Within one correlation volume, one can assume homogeneous conditions 
for the magnitude of $\phi$, but different correlation volumes will have $\phi$
magnitude varying over the entire allowed range for $\phi$ at that temperature.
If the probability of required value of $\phi$ in a single correlation domain
is $p$, then the probability of $N$ correlation domains all having the required
value of $\phi$ is $p^N$. Required value of $\phi$ for natural inflation can 
be extremely close to $\phi = 0$, with $p < 10^{-30}$ (depending on the
scale of $f$). Even if we take very liberal values of $p \simeq
0.1$, the probability of one single Hubble volume having the required value
of $\phi$ is $10^{-10^9}$ which is practically zero. We thus conclude
that the requirement of appropriate value of $\phi$ over the entire Hubble
volume is extremely fine tuned, requiring tuning much stronger than the
one inflation was intended to solve. Even if we do not take thermal initial
conditions, the problem remains as the quantum fluctuations of $\phi$ will
also be governed by an appropriate zero temperature correlation length
(which again will be determined by the potential parameters of GUT scale
leading to similar estimates). One will still need to allow that in
different correlation domains field explores the full range of values
upto vev, especially when one needs field values near the top of the
potential. 

  One may argue that one should estimate the probability of the required value
of $\phi$ over the Hubble size to be given by the decay of correlation
over the Hubble size = $e^{-H_{GUT}^{-1}/\zeta_{GUT}}$
which will be of order $e^{-1000}$ and not the absurdly small number we 
obtained above. However, this estimate is hard to justify. The correlation
length relates to the correlation of the field magnitude on the average. So,
this estimate corresponds to the {\it average} value of the field being within
the required limit over the entire Hubble volume. This is not enough for
inflation as wherever the field departs from this required range significantly
(with some other region having compensating value of the field so that the
correct average of the correlation of the field is obtained), the field will 
roll down much faster to the true vacuum. The main point is that we require 
the exact value of the field to be within the required range
over the entire Hubble volume (containing $10^9$ correlation
volumes), and not the average value of the field. (By average we mean spatial
average. If one invokes time average then the meaning of field rolling 
down the potential via field equations itself becomes ambiguous as one
will require addition of {\it noise} term.) The only way to get this
is to say that each correlation volume must have the correct value of the 
field (assuming, of course, that at least the correlation length size region
can be taken  to have the correct value of the field). This gets us back
to the earlier estimate of the probability being less than $10^{-10^9}$.

 This is a serious problem in assuming a reasonable initial condition for 
inflation. Although, the issue of initial conditions has been extensively
investigated in the literature 
\cite{berera1,berera2,robert,robertrev,east26,intl}, and inhomogeneous 
field configurations have been considered,  
generally an almost homogeneous component is taken to start the inflationary 
phase. Even for large field inflation models, there is a constraint on
acceptable variations of the field. For example, in a recent numerical 
simulation 
of full Einstein field equations coupled to scalar inflaton field, East et
al. have discussed \cite{east26} when inflation can arise from a general 
class of highly inhomogeneous initial conditions. It is then shown 
that for inflation one requires, not just the average value 
of the field, but also the fluctuation of the field in a Hubble region to 
remain within the slow roll region of the potential.
Otherwise, the field quickly rolls 
down to the minimum, being pulled down by the large field regions. This is
in accordance with the discussion above. As we mentioned, restricted
variations of the field is hard to justify from the 
picture of correlation domains where full range (allowed by the 
energy/temperature considerations) of random variations of the field occurs 
across different correlation regions.

 We will address this issue of initial conditions for inflation
using specific features of the reaction -diffusion equations. We start the
discussion of our model by briefly introducing reaction-diffusion (RD)
equations in the next section, with focus on special propagating
front solutions of RD equations.

\section{REACTION-DIFFUSION EQUATIONS IN FIELD THEORY}

Reaction-diffusion equations are studied typically in the context of 
biological systems, 
e.g. population genetics, chemical systems etc. RD equation with appropriate 
boundary condition gives traveling front with a well-defined profile. This 
profile mimics the profile of the phase boundary in a first order phase 
transition. Propagating front solutions of RD equations exist 
irrespective of the underlying phase transition dynamics  
\cite{reacdiff6,our16}.
Earlier some of us have shown the existence of propagating front 
solution in the context of chiral phase 
transition and confinement-deconfinement (C-D) transition in QCD even when the 
underlying transition is a cross-over or a continuous transition 
\cite{our16,our26,skthesis}. We have 
applied reaction-diffusion dynamics in relativistic field theory exploiting 
the fact that classical equation of motion of relativistic field in the 
presence of thermal bath has a similar form like RD equation except for 
second order time derivative term in the first case.  In the strong 
dissipation limit, the first order time derivative term in the field equation 
dominates over the second time derivative term and the resulting field 
equation of motion directly maps to reaction-diffusion equation. In the case 
of heavy ion collision experiments, a large ${\dot \phi}$ term can arise 
from the plasma expansion 
(as well as from thermal dissipation), and the required boundary conditions 
for the existence of traveling front arise from collision geometry. 
We find that even in the presence of second time derivative term, approximate
propagating front solutions of the RD equation exist, again leading to a
first order transition like dynamics. 

To demonstrate briefly how RD equation 
works for relativistic field theories, let us consider the case of a 
spontaneous chiral symmetry breaking transition for two flavor case where the 
chiral order parameter can be represented as $\Phi = (\sigma, \vec{\pi})$. 
The field equation for chiral field in chiral sigma model is,

$${\ddot \phi} - \bigtriangledown^2 \phi + \eta {\dot \phi} =
-4\lambda \phi^3 + m(T)^2\phi + h$$
\begin{equation}
m^2(T) = {m_\sigma^2 \over 2} (1 - {T^2 \over T_c^2}) .
\label{chiralfieldevolution6}
\end{equation}

Here $\Phi$ is taken to be along $\sigma$ direction only which we
represent by $\phi$ and $h$ is the explicit chiral symmetry breaking term. 
The chiral limit corresponds to $h =0$.  In the above equation,the time 
derivatives are  w.r.t the proper time $\tau$. $\eta$ is taken to be 
$1/\tau$ for Bjorken 1-d expansion of the plasma \cite{bjorken}. 
In ref.\cite{our26} constant value of $\eta$ as well as an additional 
thermal dissipation term with $\eta^\prime \propto 
T^2$ was also considered \cite{eta1}.  Here, we will establish the 
correspondence between the above field equation and RD equation by 
neglecting the explicit chiral symmetry breaking term, i.e. we take 
$h=0$. (In ref.\cite{our16,our26,skthesis} we have also presented
the case with non-zero $h$.) Let us also consider high dissipation case where 
$\ddot{\phi}$ is negligible w.r.t $\eta\dot{\phi}$. After rescaling the 
variables as ${\vec x} \rightarrow m(T) {\vec x}$, $\tau \rightarrow 
\frac{m(T)^2}{\eta}\tau$ and $\phi 
\rightarrow 2\frac{\sqrt{\lambda}}{m(T)}\phi$, the resulting equation is,

\begin{equation}
{\dot \phi} =  \bigtriangledown^2 \phi -\phi^3 + \phi .
\end{equation}

In 1-D the above equation becomes,

\begin{equation}
{\dot \phi} = \frac{d^2\phi}{dx^2}   -\phi^3 + \phi ,
\end{equation}

 which is exactly same as  the reaction diffusion equation known as 
Newell-Whitehead equation \cite{reacdiff6,hep,our16,our26,skthesis}.
The term $d^2\phi/dx^2$ can be  identified as 
the {\it diffusion term} while the other terms on the right hand side of the 
above equation  are called {\it reaction term} (representing reaction of 
members of biological species for the biological systems). 
Newell-Whitehead equation has non-trivial 
traveling front solutions  with suitable boundary conditions, namely 
$\phi = 0$ and 1 at $x \rightarrow \pm \infty$. The analytical solution with 
these boundary conditions has the form,

\begin{equation}
\phi(z) = [1+exp(z/\sqrt 2)]^{-1} ,
\end{equation}

\noindent where $z = x-v\tau$. $v$ is the velocity of the front and has the 
value $v = 3/\sqrt 2$ for this solution. In our field domain picture, the
center of the domain ($z = 0$) will correspond to one of the boundary
values of $\phi$, namely $\phi \simeq 0$, while $\phi(|z|) \rightarrow 1, 
z \rightarrow \pm \infty$. To achieve the value of the field at domain center
to be close to the boundary condition, $\phi(z=0) \simeq 0$ we need
to have large size of the domain with the following profile of the field

\begin{equation}
\phi(z) = 1 - exp((|z| - R_0)/\sqrt 2)^{-1} ,
\end{equation}

\begin{figure}[h]
\begin{center}
\includegraphics[width=0.5\textwidth]{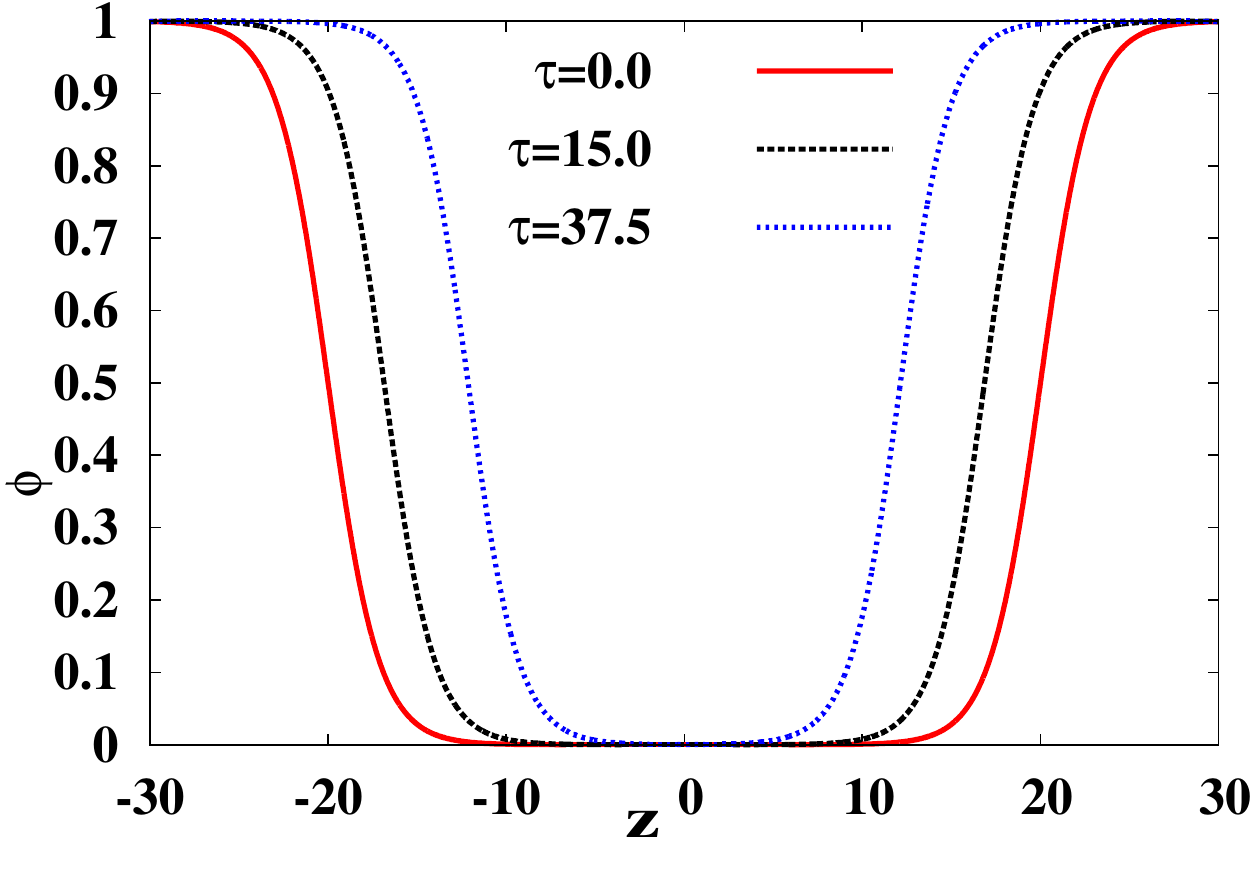}
\caption{Plot of the numerical solution for the travelling front of 
Newell-Whitehead equation at different times. Note that in this 
case symmetric initial field profile w.r.t the center is considered 
\cite{our16}}.
\label{Newell6}
\end{center}
\end{figure}

In the figure \ref{Newell6}, we have shown this field profile (with $R_0$
= 20), and the propagation of the front near the boundary of the domain
\cite{our16}. (We mention that in this particular figure in ref.\cite{our16}, 
the labels showed various quantities in dimensionful units. This is
incorrect as the plots in the corresponding figure in that  reference 
were for the above equation including domain size $R_0$, which is written in 
a dimensionless form.)  Note that the front has a shape
very similar to an interface for a first order transition and the shape
is retained as the front propagates. RD equations have several solutions, 
each with different propagating speed e.g. Newell-Whitehead equation also 
has a static solution of the form $tanh(z)$, for details, 
see \cite{our16,our26,skthesis}).  It is important to note the general 
feature of RD equation, that is, the existence of travel front solution of 
reaction-diffusion equation depends crucially on the shape of the 
underlying potential. If the potential allows for a non-zero order 
parameter in the vacuum state, along with a local maximum of the potential, 
then traveling front solution exists. The corresponding values of the 
order parameter provide the required boundary conditions for the propagating 
front solution. In ref.\cite{our16}, some of us have demonstrated how the 
dynamics 
of chiral phase transition and confinement-deconfinement phase transition can 
be dramatically changed due to these propagating solutions of the RD equation,
mimicking a first order transition even when the transition is actually
a crossover (or a second order transition).
In fact, the physics of RD equation implies that RD 
propagating front should exist whenever the potential admits a local maximum
and a minimum, with the corresponding field values providing the
required boundary conditions for the front solution. For example,
see the discussion of DCC formation discussed in ref.\cite{our26}.

\subsection{Expanding high potential energy chiral field domains in heavy-ion
collisions}
 
In ref.\cite{our26} we have extended the application of reaction-diffusion 
dynamics for the case of formation of the so called domains of disoriented 
chiral condensate (DCC) in the context of chiral phase transition.
We briefly recall that discussion here because
the discussion of RD equation in the context of  DCC formation will have a 
direct analogy with inflation. The expanding DCC domain will be exactly
analogous to the inflating {\it field domain}. Further, the chiral sigma
model potential has exactly the same features as the natural inflation model.

DCC is an extended region where the chiral field is misaligned from the 
true vacuum, hence has high potential energy. This topic has been extensively 
discussed in the literature in the context of explaining the anomalous result 
in Centauro events in cosmic ray experiments \cite{centa6}. A large 
DCC domain can give rise to the coherent emission of pions when the chiral
field rolls down to the true vacuum (quite like the roll down of the
inflaton field at the end of inflation) \cite{bjorkendcc6}. The original
proposal had difficulties as expected DCC domain was small  which will
not give clean signals of coherent pions. Some of us have shown in 
\cite{our26} that in high multiplicity proton-proton collisions,
initially small DCC domains can have correct profile as needed for slowly
moving propagating fronts of RD equations, thereby strongly delaying
the roll down of the chiral field to the true vacuum (again, despite the
fact that there is no metastable vacuum). Expanding plasma then stretches
this domain where the chiral field is still stuck near the top of
the potential (strongly disoriented) leading to a large DCC domain. It
is now clear that this is exactly the situation one would like to
achieve for the inflaton field inside a Hubble volume, and as we will
see below this indeed can be achieved. Below we will provide a brief
summary of DCC formation via RD equation from ref.\cite{our26}.

 The formalism of DCC formation is discussed in the framework of linear 
sigma model. Physics of chiral symmetry 
breaking is an important ingredient of chiral sigma model. When the 
temperature of the system drops below the critical temperature, the chiral 
field ($\Phi = (\sigma, {\vec \pi}$) as discussed for Eq.(1)) picks up 
random directions in the vacuum manifold in different regions 
in the physical space. Lagrangian density of this model is given by,

\begin{equation}
L = {1 \over 2} \partial_\mu \Phi \partial^\mu \Phi
- V(\Phi,T) ,
\end{equation}

where the finite temperature effective potential $V(\Phi,T)$ at 
one loop order is given by,

\begin{equation}
V = \frac{m_\sigma^2}{4} \left({T^2 \over T_c^2} - 1\right) 
|\Phi|^2 + \lambda |\Phi|^4 .
- h \sigma
\end{equation}

\begin{figure}[h]
\begin{center}
\includegraphics[width=0.5\textwidth]{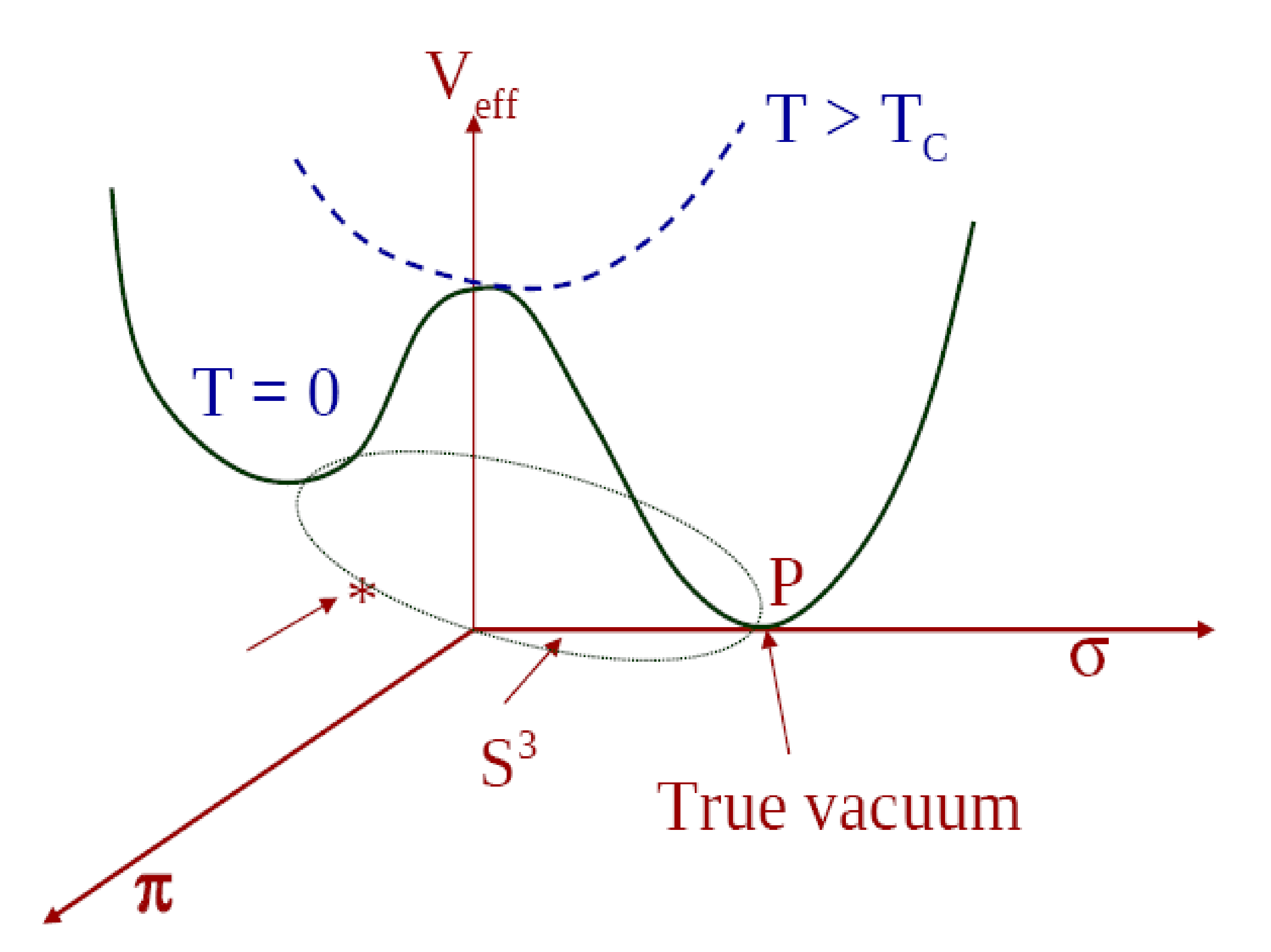}
\caption{Effective potential for the chiral field $\Phi$. P denotes the 
true vacuum on the (approximately degenerate) vacuum manifold while $\star$ 
marks the value of the chiral field inside a DCC domain which is 
disoriented from the true vacuum direction \cite{our26}.}
\label{dccfigure6}
\end{center}
\end{figure}

 We take parameter values as in ref. \cite{our26}. In the absence of 
explicit chiral symmetry breaking (i.e. $h = 0$), the 
vacuum manifold is $S^3$ where all the points on vacuum
manifold are equally likely. But in the presence of explicit chiral symmetry 
breaking term ($h \ne 0$), there is a unique 
vacuum state. Shape of the chiral effective potential in the presence 
of explicit symmetry breaking is shown in the figure \ref{dccfigure6}\,.

In a rapidly cooling system, 
one expects that the chiral field will take some arbitrary value in the vacuum 
manifold within a correlation domain. If this value differs from the true 
vacuum then one gets DCC domains. Subsequently, the field will roll down to the 
true vacuum, leading to coherent pion emission. To study the roll down of 
the chiral field we have used field equations from Lagrangian in Eqn.(6)
(which, as we have argued above, leads to reaction-diffusion dynamics).
Note that, as we mentioned above, we use a 
second time derivative term as well, and still the propagating front
solution exists (apart from some oscillations arising from the second
time derivative term).  We have considered the case of  maximal disorientation 
where field value at the center of the domain is opposite to 
the true vacuum on the vacuum manifold. (We also considered field
value slightly away from the saddle point, and our results remain almost
unchanged, see \cite{our26} for a discussion of this point.) 
Outside the domain, field takes true vacuum value. With this initial 
condition, we used field equation appropriate for the chiral phase 
transition. It is important to note that this 
boundary condition does not guarantee the propagating front solution. For RD 
equations, the corresponding boundary conditions are set for the local maximum 
of the potential. But in the case of DCC, the boundary condition is set for 
the saddle point. Interestingly saddle point boundary condition also works 
because the requirement of local maximum of the potential is satisfied
at that point in the angular direction.  Thus boundary 
conditions with saddle point and the true vacuum seem appropriate, and
indeed these give rise to traveling 
front solution. If one starts with the field configuration such that at 
boundary the field value is close to the saddle point, then in that case 
the field starts rolling down slowly towards the true vacuum, though
slowly propagating front still exists \cite{our26}. 

Here we highlight the major achievements of 
using RD equation. Normally one would 
expect roll down of the field from the saddle point to the true vacuum in a 
timescale of the order of fm.  However, because of the reaction-diffusion 
dynamics, front solution delays this roll down. The field retains its value 
close to the saddle point in a significant region for long duration of time. 
During the slow rolling of the field due to the reaction-diffusion dynamics, 
plasma continues to expand, therefore the region where the chiral field was 
initially disoriented stretches to a larger region. This results in a DCC 
domain which is expanding and getting bigger without the chiral field in the 
interior rolling down towards the true vacuum. Extra potential energy coming
from the kinetic energy of plasma expansion (just like the situation of 
negative pressure where expansion requires input of energy).  As an example, 
if one starts with DCC domain of size of order 2-3 fm initially, then due 
to RD equation solution, the size of the DCC domain can grow to several 
times larger than the original one, and within this large domain the field 
value will still remain close to the saddle point. 
This stretching of the field domain is exactly
what we need for the inflation case. The shape of the potential for
natural inflation is exactly similar to the linear sigma model case for
the chiral field rolling down the valley of the potential. Only 
difference between the two potentials is the relevant energy scales. 
We now turn to the discussion for the case of inflation.

\section{REACTION-DIFFUSION EQUATION FOR NATURAL INFLATION}

In this section, we will use propagating front solutions of the RD equation, 
as used above for DCC formation, to address the issue of initial 
condition in case of Natural inflation model. We will not discuss the issue of 
the viability of the inflation model in this paper. Any model of inflation 
has to address the issue of initial conditions 
for the field. Also, our model can be extended 
to other models of inflation. We take the PNGB potential of the form,

\begin{equation}
V(\phi)=\Lambda^4(1+cos(\phi/f))
\label{naturalpotentail6}
\end{equation}

\begin{figure}[h]
\begin{center}
\includegraphics[width=0.5\textwidth]{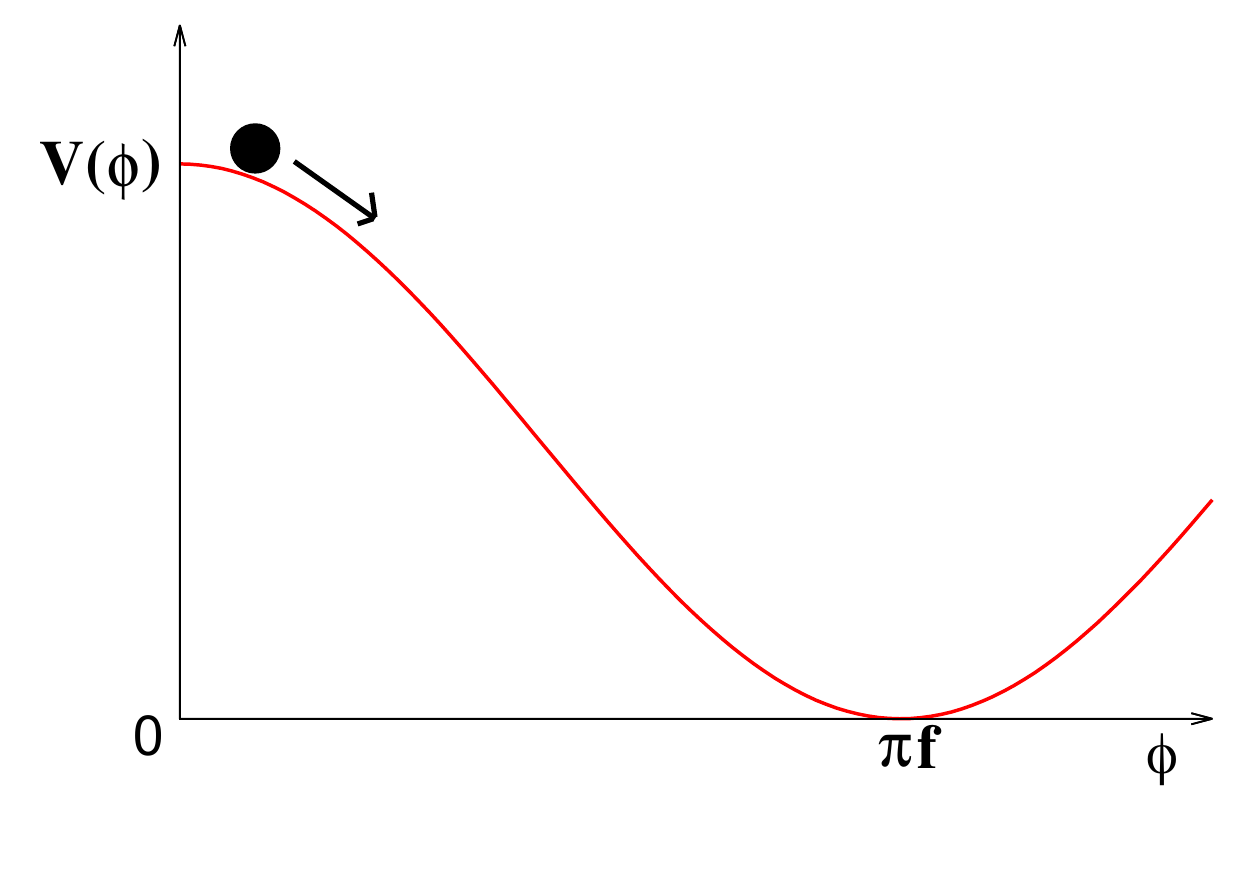}
\caption{ Plot of the potential for natural inflation. Note, $\phi$ 
here corresponds to the angular variable in Fig.\ref{dccfigure6}\, while 
the top of the potential corresponds to the height of the saddle point 
opposite to the true vacuum (P).}
\label{fig46}
\end{center}
\end{figure}

The shape of this is given in Fig.\ref{fig46}. To relate to the chiral model 
case of figure \ref{dccfigure6}\,, \,$\phi$ corresponds to
the angular variable in Fig.2, and $\Lambda$ is related to the explicit 
symmetry breaking term $h\sigma$ in Eq.(7). The height of the 
potential in Eq. \ref{naturalpotentail6} is $2\Lambda^4$. The potential has 
a unique minimum at $\phi = \pi f$ (which corresponds to the true vacuum
for the chiral potential in Fig.\ref{dccfigure6}). $\phi = 0$ is the local 
maximum of the potential and corresponds to the saddle point in 
Fig.\ref{dccfigure6}. The periodicity of $\phi$ is $2\pi f$. We have 
taken $f = m_{pl}$ and $\Lambda = 10^{15}$ GeV  which satisfies the constraints 
coming from slow roll condition, the requirement of enough inflation and 
observational constraint on the density contrast at the surface of the last 
scattering \cite{natural61}. 
Though it is important to realize that these constraints were
derived for a homogeneous field configuration which has value close
to $\phi = 0$ over several Hubble volumes. Thus, in the field evolution
$\bigtriangledown^2 \phi$ term was ignored.  For our case this term plays
crucial role in stabilizing the non-trivial profile of the propagating
front. So, appropriate constraints in our case will arise by comparing 
vacuum energy to the kinetic energy as well as the gradient energy
of the field configuration. However, now all these three contributions
depend on the exact field profile in the domain, as well as exact
dynamics of the field. So, kinetic energy  does not arise only from the
roll down of the field from near the top of the potential, but also
from the motion of the propagating front which is well inside the
Hubble volume (the entire field domain being smaller than the
initial Hubble volume at the GUT scale). Further, there is gradient energy
contribution which changes as the front evolves. So, no analytic expressions
can be written down for these different contributions unless one
finds an exact solution for the propagating front for the field
equation in expanding universe. It certainly will be very interesting
to find such exact solutions of these {\it generalized} RD equations. 
We only use the parameter values for the natural inflation
as in the literature for the conventional case, and show that inflation
is possible via RD equation fronts. It is quite possible that quite
different parameter choices may also provide inflation (with correct
fluctuations) when RD equation solutions are used.

 Since the potential is symmetric about its minimum, we assume that inflation 
begins with the value of the field inside a domain, $\phi = \phi_1$, 
$0<\phi_1/f<\pi$. For sufficient inflation, $\phi$ should satisfy  the 
condition  $0\leq \phi_1\leq \phi_1^{max}$, e.g if 
one takes $f=m_{pl}$ then $\phi_1^{max}/f$ is about $0.2\pi$. 
We assume that the Hubble volume at the GUT scale is filled with radiation 
energy (with typical number of degrees of freedom as appropriate for
GUT scale, we take it to be 100), in addition to that Hubble volume contains 
several field domains where field profile takes non-trivial shape.
As the potential energy for the inflaton field (in this natural inflation
model with PNGB field) arises at the GUT scale, it is reasonable to assume
that inside these domains the field can have high potential energy, while 
outside the domain the field lies in the true vacuum. Thus typical profile
in each field domain will be the field rising up towards the top of the 
potential inside the domains, while smoothly changing to the true vacuum
value outside the domains, as was discussed above.  
In different correlation domains $\phi_1$ will be
expected to be randomly distributed between $0$ and $\pi f$. In all the domains
where the field inside the field domain is not very close to zero, the
field will very quickly roll down to the true vacuum. 

We consider a single domain where $\phi_1$ is very close to zero inside the 
domain (changing smoothly to vev at the boundary of the domain),
and this is
the only domain which will survive with nontrivial field profile as the 
Universe evolves.  In the standard picture if we do not invoke 
reaction-diffusion dynamics then one expects that in the time scale 
given by the shape of the potential, the field $\phi$ should rapidly roll 
down to the true 
vacuum with zero potential energy. But in the presence of reaction-diffusion 
dynamics situation is very different.  We will show that traveling front 
solution of reaction-diffusion equation delays the roll-down of $\phi$.
The field rolls down very slowly in the interior of the domain, and the
size of the domain also shrinks extremely slowly (in comoving coordinates).
End result is that vacuum energy starts dominating the average energy
density in that Hubble volume. The Hubble region then enters inflationary
stage. Eventually one will achieve homogeneous field configuration over the 
entire Hubble volume starting from a very general inhomogeneous initial 
condition.  This will be the stage of beginning of proper inflationary
stage of the Universe.
Duration of inflation will depend on the roll down of the field
at the center of the domain, which will depend on the exact profile
of the field in that domain. Important thing to note here is that we do 
not take the field to be homogeneous over several Hubble volumes (or even
one Hubble volume). Rather, we take the field to be roughly uniform over 
a smaller region.  This is the most important difference between 
the traditional models of inflation and our model. 

Once the initial conditions are fixed, now we can focus on 
the dynamics of the inflaton field. With the above choice of potential, 
the  equations governing the dynamics of  inflaton field become, 

\begin{equation}
\ddot{\phi}-\frac{\nabla^2\phi}{a^2}+3H\dot{\phi} -
\frac{\Lambda^4}{f}sin(\phi/f)=0
\label{fieldevolution6}
\end{equation}

\begin{equation}
H^2=\left(\frac{\dot{a}}{a}\right)^2=\frac{8\pi}{3m_p^2}(\rho_{\phi}+\rho_{rad})
\label{scalefactorevolution6}
\end{equation}

\begin{equation}
\rho_{\phi}=\frac{1}{2}\dot{\phi}^2+\frac{(\nabla \phi)^2}{2a^2}+
\Lambda^4(1+cos(\phi/f))
\label{energydensity6}
\end{equation}

\begin{equation}
\rho_{rad}\propto a^{-4}
\label{radiationenergydensity6}
\end{equation}
 
where Eq. \ref{fieldevolution6} is the classical equation of motion of the inflaton 
field in FRW background. Eq. \ref{scalefactorevolution6} gives the evolution of the scale factor $a$. $\rho_{\phi}$ in Eq. \ref{energydensity6} is the energy density associated with the 
scalar field and Eq. \ref{radiationenergydensity6} gives the standard evolution of the radiation
energy density. We 
rescale the variables as follows, ${\vec x} \Lambda \rightarrow {\vec x}$, 
$t\Lambda \rightarrow t$, $\phi/f\rightarrow \phi$, $H/\Lambda\rightarrow H$. 
The resulting equations are,

\begin{equation}
\ddot{\phi} + 3H\dot{\phi} + \frac{\Lambda^2}{f^2}sin(\phi) - 
\frac{(\nabla^2 \phi)}{a^2} = 0
\label{inflatonevo6}
\end{equation}

$$H^2 = \left(\frac{\dot{a}}{a}\right)^2 = 
\frac{8\pi}{3}\frac{f^2}{m_{pl}^2}\bigg(\frac{1}{2} \dot{\phi}^2 + 
\frac{1}{2}\bigg(\frac{\nabla\phi}{a}\bigg)^2 + $$ 
\begin{equation}
\frac{\Lambda^2}{f^2}\bigg(1+ cos(\phi)\bigg) +
\frac{\rho_{rad,0}}{f^2\Lambda^2}\bigg(\frac{a_0}{a(t)}\bigg)^4\bigg)
\label{scaleevolutiondetail6}
\end{equation}

where we have taken the value of $\rho_{rad,0} = g^* \Lambda^4$ i.e. radiation 
energy density at GUT scale with number of degrees of freedom $g^* = 100$. 
Recall, in our model we have taken $\Lambda = 10^{15}$GeV, and  $f = m_{pl}$. 
If we neglect the $\ddot \phi$ term in Eq. \ref{inflatonevo6} and take $H$ 
to be constant then Eq. \ref{inflatonevo6} becomes a reaction-diffusion 
equation (with time dependent
diffusion constant). The potential term here (the reaction term for
the RD equation) is new and we have not seen any standard RD equation
with this form of the reaction term. From the general analysis of
RD equations, we expect this equation also to have well defined propagating
front solutions for appropriate boundary conditions. Again, as for the
chiral field case, we expect approximate solutions to survive even in the
presence of second time derivative term, as well as with time dependent
$H$, as in Eq.\ref{inflatonevo6}\,. 

To solve this set of differential equations we 
need the initial field profile which satisfies proper boundary condition. 
We have taken $tanh$ profile of the field interpolating between minimum of the 
potential at $\phi = \pi f$ and the maximum of the potential at 
$\phi = 0$. We have taken this profile for 1-D (planar propagating front)
as well for 3D case (spherically symmetric propagating front). 
In our earlier works \cite{our16,our26,skthesis} when we have investigated 
reaction-diffusion dynamics, we have also taken exact solution of the 
corresponding RD equation as initial profile which has zero velocity. 
In the present case we take $tanh$ profile  as an example (which has
sufficiently rapid fall off behavior which helps us in doing simulation
with limited size of lattice). Exact solutions for this {\it generalized}
RD equation are not known. Further, it is very important to mention that 
even if one starts with a different profile 
interpolating between correct boundary values, very quickly the profile
changes to the appropriate profile corresponding to the RD equation under
consideration. In fact even a profile consisting of linear segments (with
correct boundary conditions) rapidly becomes smooth and acquires the
correct shape as appropriate for the solution of the RD equation, 
see, discussion in ref.\cite{our16,our26,skthesis}). (Thus, the use of
$tanh$ profile has nothing to do with any assumption of thermal equilibrium
for the inflaton field. Indeed, we discuss a case in section V.C where 
field equations do not have any thermal dissipation present).  So our results 
for the propagating front are not sensitive to the initial profile. However,
the choice of the profile also determines the value of the field at the center
of the domain, which eventually will determine the full duration of inflation.
The issue of dependence of duration of inflation on the exact profile of
the field requires long time simulations, and will be addressed in a future
publication. As we mentioned above, it is not easy to determine
the roll down of the field at the center
of the domain simply from field equations by
neglecting the space dependence of field as $\nabla^2 \phi$ terms
plays most crucial role in the motion of the propagating front (hence
roll down of the field).   

  As we mentioned earlier, $\dot \phi$ term plays crucial role in the
propagating front solutions of the RD equations. For this purpose we have
also considered a thermal dissipation term in Eq.\ref{inflatonevo6} along 
with the $H \dot \phi$ term. With this, the equation of motion for 
$\phi$ becomes

\begin{equation}
\ddot{\phi} + (3H + \eta) \dot{\phi} + \frac{\Lambda^2}{f^2}sin(\phi) - 
\frac{(\nabla^2 \phi)}{a^2} = 0
\label{eomphi6}
\end{equation}

We have taken thermal dissipation term to be proportional to $T$. In
the literature different models for thermal dissipation have been discussed
leading to $T$ as well as $T^2$ dependence \cite{eta1,eta2}. 
We take $T$ dependence for the
following reason. First this leads to a large contribution from thermal
dissipation compared to Hubble term which gives better results for RD equation
fronts. Though we will also present results without any thermal dissipation
term, i.e. with only $3H \dot \phi$ term as in Eq.\ref{inflatonevo6}, but 
that requires larger domain size (still smaller than the GUT Hubble size).
One can justify $\eta \sim T$ dependence compared to $T^2$ dependence as
with $T^2$ dependence the $H\dot \phi$ term and $\eta \dot \phi$ term will
always have the same relative strengths in a radiation dominated universe
(with both $H$ and $T^2  \sim 1/t$). This is not reasonable as  thermal
dissipation arises from particle interactions and we expect Hubble expansion
effects to become less relevant for later times for micro physics of
particle interactions. Thus, we take $\eta$ to be of same magnitude as
3$H$ at the Planck stage corresponding to the case when the Universe is
in thermal Equilibrium at the Planck stage. Subsequently we take $\eta$ to 
decrease as proportional to $T$.  Note, this is only for estimate of $\eta$, 
we never use this equilibrium Planck stage in our model. In any case, use 
of thermal equilibrium will raise issues of strong coupling
for the inflaton. This is an appropriate place to clarify that our aims 
are rather modest in this paper, just to show how an inhomogeneous
domain can enter inflation. By no means we aim to address many of
the important problems of inflationary models.  With this clarification, 
we may note that it is possible that due to a very different dynamics of 
inflaton field in this case where roll
down of the field is delayed due to special nature of the propagating
front, such constraints may be relaxed. Certainly it is not guaranteed
and one needs to explore these issues in detail in the context of this
particular Reaction-Diffusion front mediated field evolution where
field gradients play crucial role.

With above discussion, in scaled coordinates $\eta$ is given by

\begin{equation}
\eta(T) = {T \over T_{Pl}} 3H_{Pl} = 3 {T_{Pl} \over T_{GUT}} {H(T_{GUT})
\over a}
\end{equation}

with the scale factor $a$ taken to be 1 at the GUT scale.
Thus at the GUT scale of 10$^{15}$ GeV, $\eta$ becomes about
10$^4$ times larger than the GUT scale Hubble constant. However, effective
Hubble constant for our case is much larger due to large contribution of
field gradient energy. With that large value of $H$, thermal dissipation
is found to be larger than  $3H$ by a factor of about 500 at the GUT scale. 
Subsequently, thermal dissipation becomes much larger as temperature 
decreases. This thermal dissipation term is important for us only until  the
Hubble volume  enters inflation (more conservatively when the domain
exits the Hubble volume). We find that by that time temperature only
decreases by about 4-5 orders of magnitude (from the GUT value),
so it is reasonable to take the standard picture of thermal dissipation.
Subsequently, when inflation is established then even if thermal dissipation
becomes ineffective, it is of no consequence.

\section{NUMERICAL RESULTS}

We will consider the domain to be spherically symmetric such that
the field in the interior of the domain takes the value near the top of
the potential, while it smoothly changes to the true vacuum value at the
boundary of the domain. In this sense, this domain is something like
an {\it inverted} bubble profile for a first order transition (as
the interior of the domain has high potential energy, note though that
the potential here corresponds to a second order transition). For the
evolution of the profile of the field in this domain, first we will consider
a 1-D solution, where $\nabla^2\phi = \frac{d^2\phi}{dz^2}$. For large 
domains this should be a good approximation as the domain boundary will be
approximately planar. 1-D profile is used primarily due to boundary conditions
for the numerical solution as we will discuss below.
We will later also give solutions of 3-D equation. We also include
thermal dissipation as discussed above. Later we will present results also
for the case without any thermal dissipation.  All the length and time scales 
are expressed in the units of $\Lambda^{-1}$ and $\phi$ is in the units 
of $f$. 

\subsection{1-D field profile}

We take the propagating front to be planar, so that the resulting
spatial derivative becomes one-dimensional. The field equation 
Eq. \ref{eomphi6} becomes:

\begin{equation}
\ddot{\phi} + (3H + \eta) \dot{\phi} + \frac{\Lambda^2}{f^2}sin(\phi) - 
{1 \over a^2}{d^2\phi \over dz^2} = 0.
\label{1dfieldevolution6}
\end{equation}

 For large domains this should be a good approximation.
For 1-D equation we have taken symmetric initial profile w.r.t to the center 
of the domain which is taken to be $z = 0$.  The field profile from the 
center in positive z direction is taken as

\begin{equation}
\phi(z) =  {\pi \over 2} [1 +  tanh((z-z_0)/d)].
\label{tanh6}
\end{equation}

Here, $z_0$ characterizes the size of the domain where field is very close
to 0, and $d$ gives initial thickness of the region in which field changes
to the true vacuum value ($\pi$) (recall, $\phi$ is in units of $f$ in 
Eq.\ref{eomphi6}).
The reason we evolve $\phi$ in the full domain, even though the field is
symmetric about the domain center, is because for numerical evolution
we need to fix boundary conditions. If we fix $\phi$ in the domain center 
to be at the top of the potential ($\phi = 0$) then it will raise concern 
whether the evolution of the front, and any possible roll down of the field, 
is crucially affected due to this fixed boundary condition (even though 
for large value of $z_0$,
$\phi$ at domain center should not matter). To avoid such concerns we fix
the field at the two ends of the domain to lie at the true vacuum value,
so that the field is free to roll down in the center. This is also the reason
we work with 1-D solution first. For 3-D case, with the radial profile
of $\phi$, there is no option but to fix $\phi = 0$ at $r = 0$. However, as
we have already found slowly propagating well defined fronts for the
1-D case, we are confident that the fixed boundary condition at $r = 0$
for the 3-D case has insignificant effect on the evolution of the field 
profile. (This is for the time scale for which we have carried out the 
simulation. Eventually the field will roll down depending on the relative 
size of the domain and the Hubble size, and at that stage the role of 
fixed boundary condition at $r = 0$ will become important.) 

\begin{figure}[h]
\begin{center}
\includegraphics[width=0.5\textwidth]{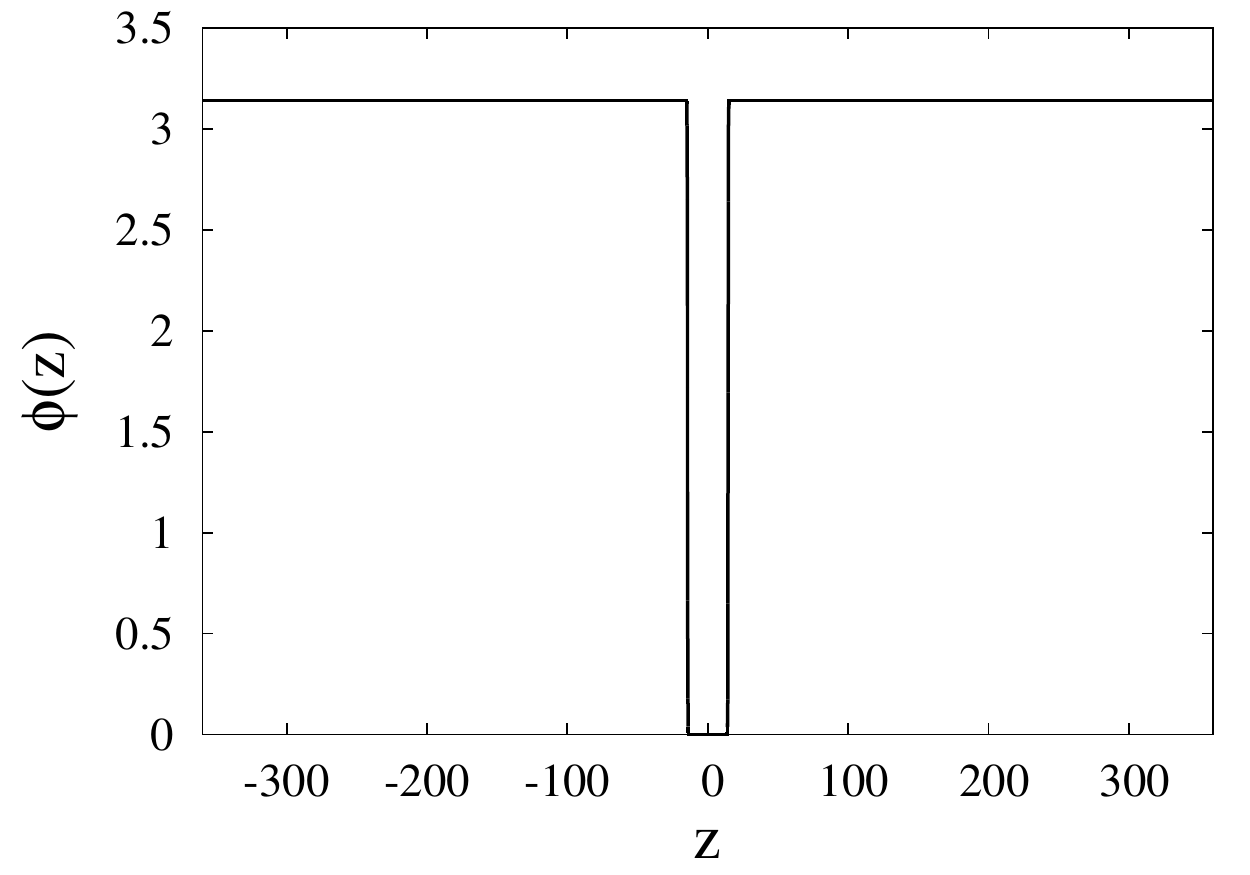}
\caption{ Profile of $\phi$ for the initial {\it field domain}. The diameter
of the domain is about 30 (everywhere, time and lengths are in units of 
$\Lambda^{-1}$).  This is also the physical size of the domain as the 
scale factor $a$ is taken to be 1 initially at the GUT scale.
Note that $H_{GUT}^{-1} \simeq 723$ for GUT scale of $10^{15}$ GeV. 
This entire $H_{GUT}^{-1}$ region is shown in the figure
to illustrate that the field domain is smaller initially than the causal 
horizon.  However, the
field domain has very large contributions from the gradient of the field
due to non-trivial spatial profile of $\phi$. Including this contribution
for the region containing the field domain, the actual value of $H^{-1}
\simeq 33$, which is still larger than the size of the field domain.}

\label{fig56}
\end{center}

\end{figure}

The field profile at the initial stage (GUT scale) is shown in 
Fig \ref{fig56}. The diameter of this {\it field domain} is 
about 30 (everywhere, time and lengths are in units of $\Lambda^{-1}$).
This is also the physical size of the domain as the scale factor $a$ is 
taken to be 1 initially at the GUT scale. $H_{GUT}^{-1} \simeq 723$ for 
GUT scale of $10^{15}$ GeV. This entire $H_{GUT}^{-1}$ region is shown
in Fig. \ref{fig56} to illustrate that this field domain is smaller 
initially than the causal horizon. However, the energy of the field domain 
has very large contributions from the gradient of the field
due to non-trivial spatial profile of $\phi$. Including this contribution
for the region containing the field domain, the actual value of $H^{-1}
\simeq 33$, which is still larger than the size of the field domain we
have taken. We again mention that we are working with a strong assumption
of using FRW metric with such highly inhomogeneous energy density 
distribution. We expect that a complete solution of the problem will 
still not affect the roll down of the field being governed by (possibly 
a generalized) RD equation front, though it will certainly affect the local 
expansion rate.

 We carry out the simulation by simultaneously solving 
Eqn.14 (for 1-D) and Eqn.17. So, for the initial 
field profile of Fig.\ref{fig56}, we calculate total field energy,
and by taking the GUT scale Hubble size as the region of interest, we
calculate field energy density. We add to this the radiation energy
density at the GUT scale ($\sim g^* \Lambda^4$ with $g^* = 100$) and
then calculate resulting Hubble constant for this total energy density.
Due to very large contribution of field energy, the final Hubble constant
is much larger than the GUT scale Hubble constant. We find that the
net value of $H^{-1}$ is about 33 which is smaller than $H_{GUT}^{-1}
\sim 723$. With this new value of $H$ the field equation 
(Eq.\ref{1dfieldevolution6}) is solved for
the evolution of the field profile. The evolution of field configuration
changes relative proportions of vacuum energy contribution, and the gradient
and kinetic energy of field, as well as radiation energy, and new Hubble
parameter is calculated at each time step with this changed energy density.
We have evolved field configuration up to $t = 110600$ by which time the
Hubble parameter becomes almost constant due to complete dominance
of the vacuum energy density of the field profile over all other energy
contributions. 

\begin{figure}[h]
\begin{center}
\includegraphics[width=0.5\textwidth]{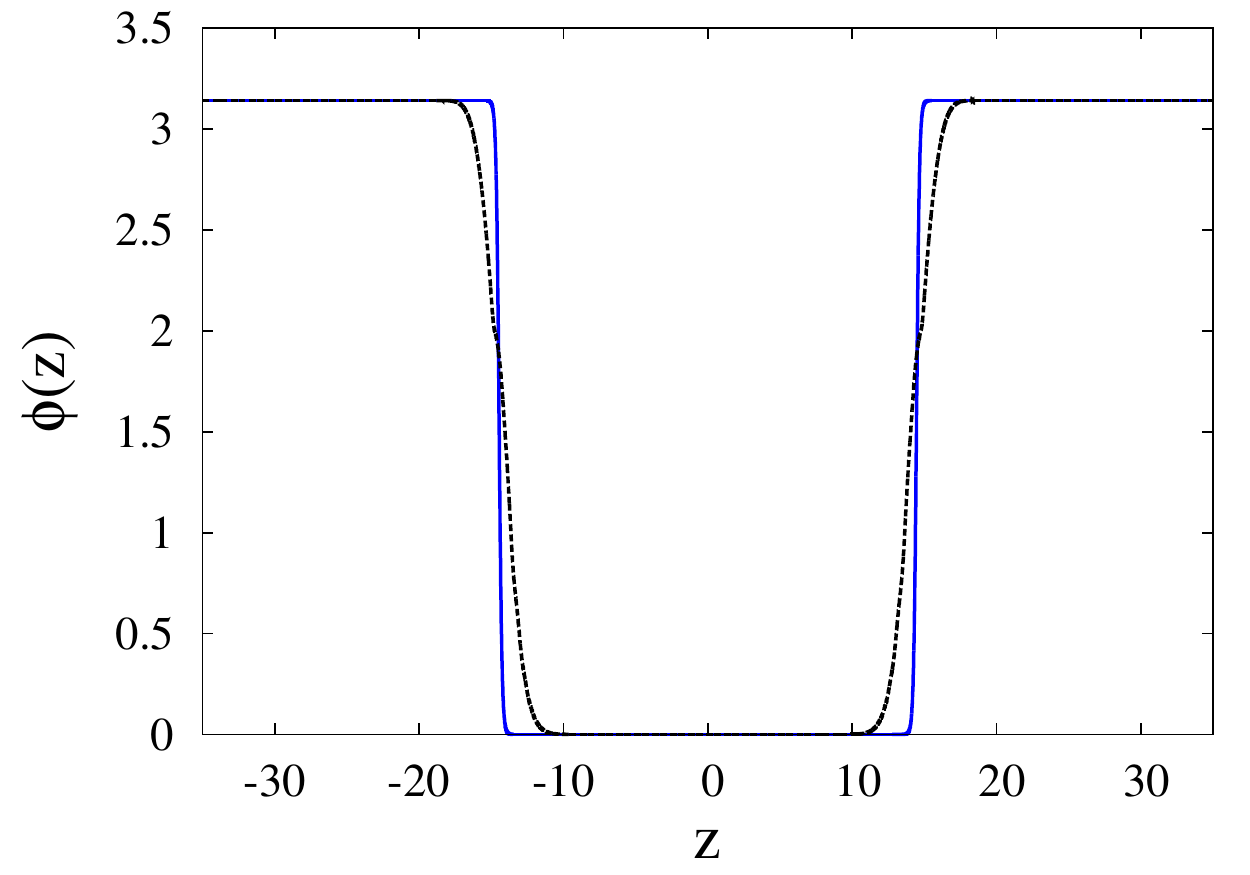}
\caption{Final plot of $\phi$ in comoving coordinates at $t \simeq 110600$ is 
shown by the black (dashed) curve. The blue (solid) curve shows the initial 
profile of the field. The initial size of the actual Hubble region $H^{-1}$,
incorporating field energy contributions, is about 33, which is still
larger than the initial domain size.}
\label{fig66}
\end{center}
\end{figure}

 Fig.\ref{fig66} shows the final profile of $\phi$ at $t = 110600$ and
 the initial profile in the comoving coordinates. Note that 
the shape of the propagating front has changed, primarily to adopt to the 
correct solution of this generalized RD equation (Eq.\ref{1dfieldevolution6}). 
The shrinking of the
domain in comoving coordinates is insignificant and expansion of the Universe
leads to stretching of the domain in physical coordinates.
The initial size of the actual Hubble region $H^{-1}$,
incorporating field energy contributions, is about 33, which is still
larger than the initial domain size.

\begin{figure}[h]
\begin{center}
\includegraphics[width=0.5\textwidth]{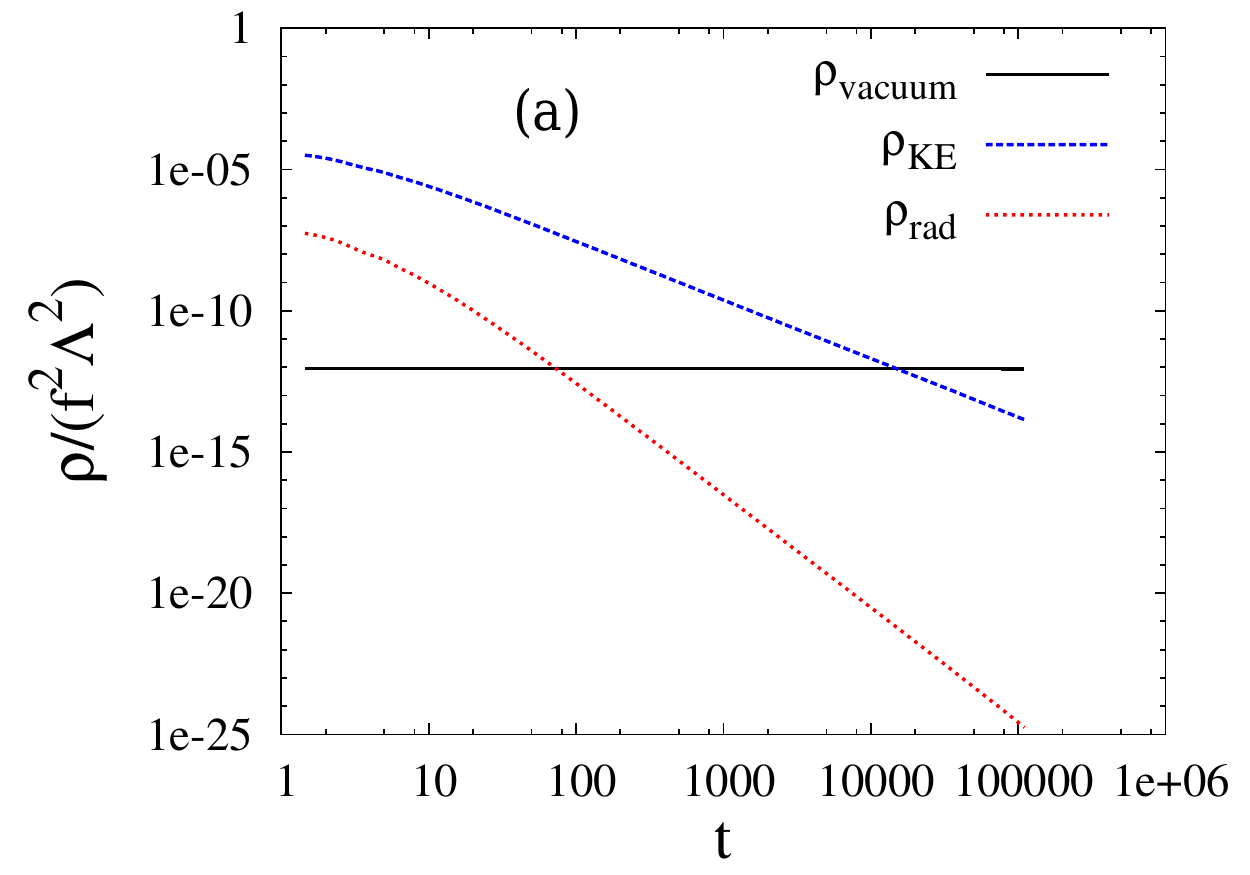}
\includegraphics[width=0.5\textwidth]{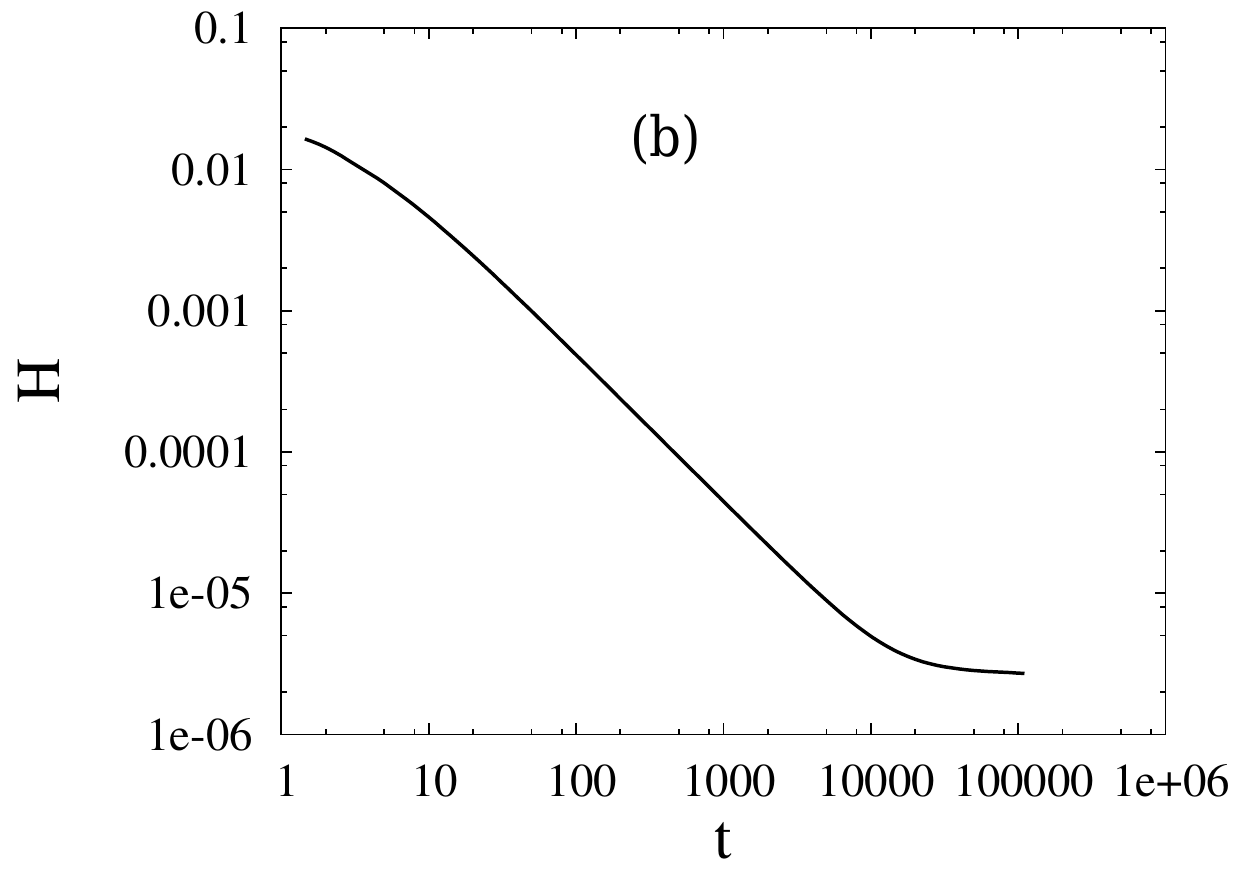}
\caption{(a) Log-Log plot of the evolution of different components of 
the energy density
as a function of time. Solid (black), dotted (red), dashed (blue)
curves show the contributions of the field vacuum energy, the field
kinetic energy + gradient energy (denoted by $\rho_{KE}$), and the radiation
energy density respectively. The vacuum energy becomes most dominant 
component at $ t \simeq 14923$. (b) Log-Log plot of the evolution of 
the Hubble parameter $H$. During early stages $H$ changes non-trivially,
subsequently, $H$ evolves as a power law, eventually turning over to
a constant value of $H$ signaling the beginning of the inflationary
phase. In (b), this happens around at $t \simeq 15000$ which 
coincides with the stage of the dominance of the vacuum energy in (a).}
\label{fig76}
\end{center}
\end{figure}

Fig.\ref{fig76} is the log-log plot of the evolution of energy density as well
as of the Hubble parameter. Fig.\ref{fig76}\,a shows the evolution of 
different components of the energy density as a function of time. Note, 
initially, the most dominant component is the kinetic energy + gradient 
energy part of the field (which actually leads to much
larger value of $H$ than what is expected at the GUT scale). With time,
all other components decrease except the vacuum energy density which
remains essentially constant. Minor changes in vacuum energy density
occur due to adjustment of the field configuration with changed $H$
(through the $\dot \phi$ term) during the evolution of the propagating
front. We find that the vacuum energy becomes most dominant component at
$ t \simeq 14923$. Fig.\ref{fig76}\,b shows the evolution of the Hubble 
parameter $H$ (Eqn.(14)).
During early stages $H$ changes non-trivially due to the adjustment of
propagating front to correct solution of the RD equation (also note,
time is taken as $t = 0$ at the GUT scale, so there is a constant
$t_{GUT}$ in writing a relationship between the scale factor $a$ and $t$).
Subsequently, $H$ evolves as a power law, eventually turning over to
a constant value of $H$ signaling the beginning of the inflationary
phase. In Fig. \ref{fig76}\,b, this beginning of the inflationary phase
happens around at $t \simeq 15000$ which 
coincides with the stage of the dominance 
of the vacuum energy in Fig.\ref{fig76}\,a.

\begin{figure}[h]
\begin{center}
\includegraphics[width=0.5\textwidth]{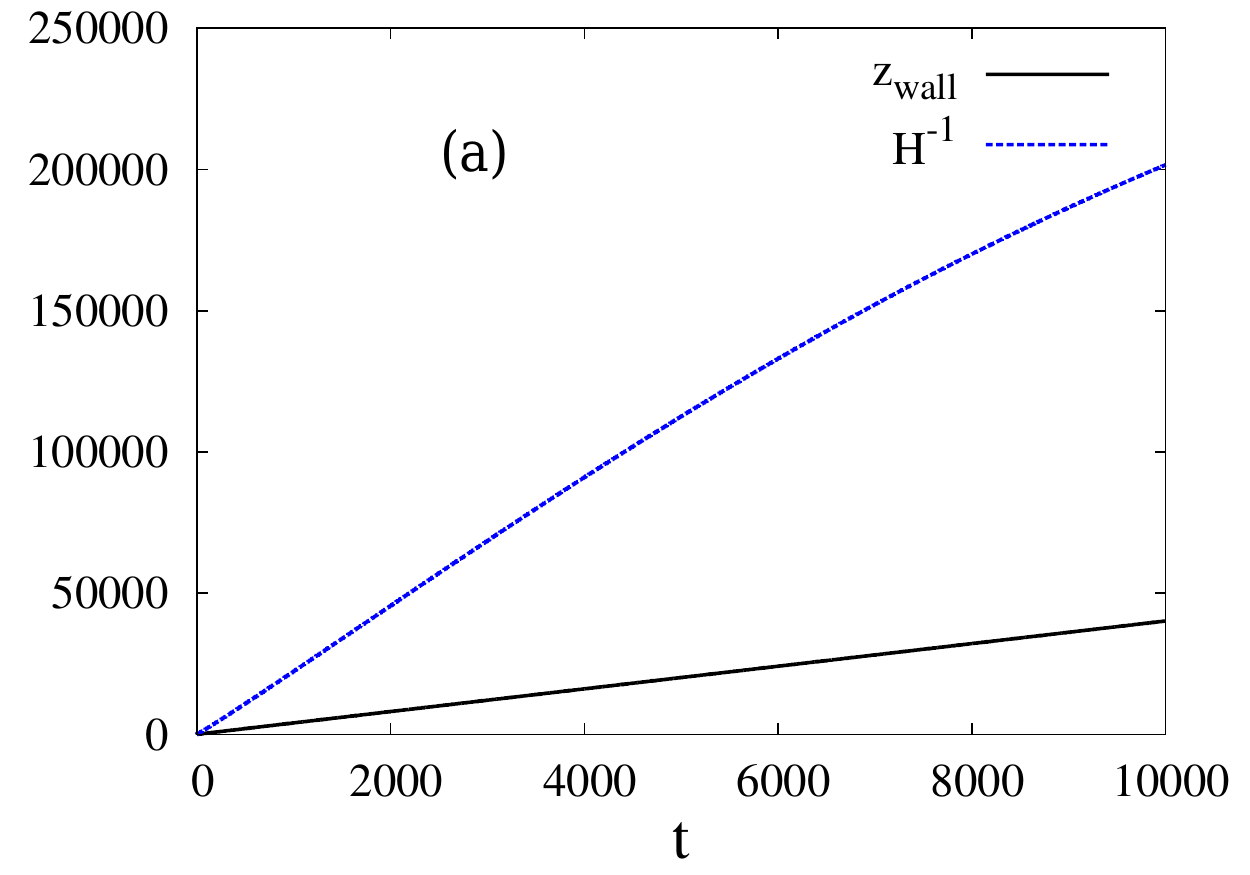}
\includegraphics[width=0.5\textwidth]{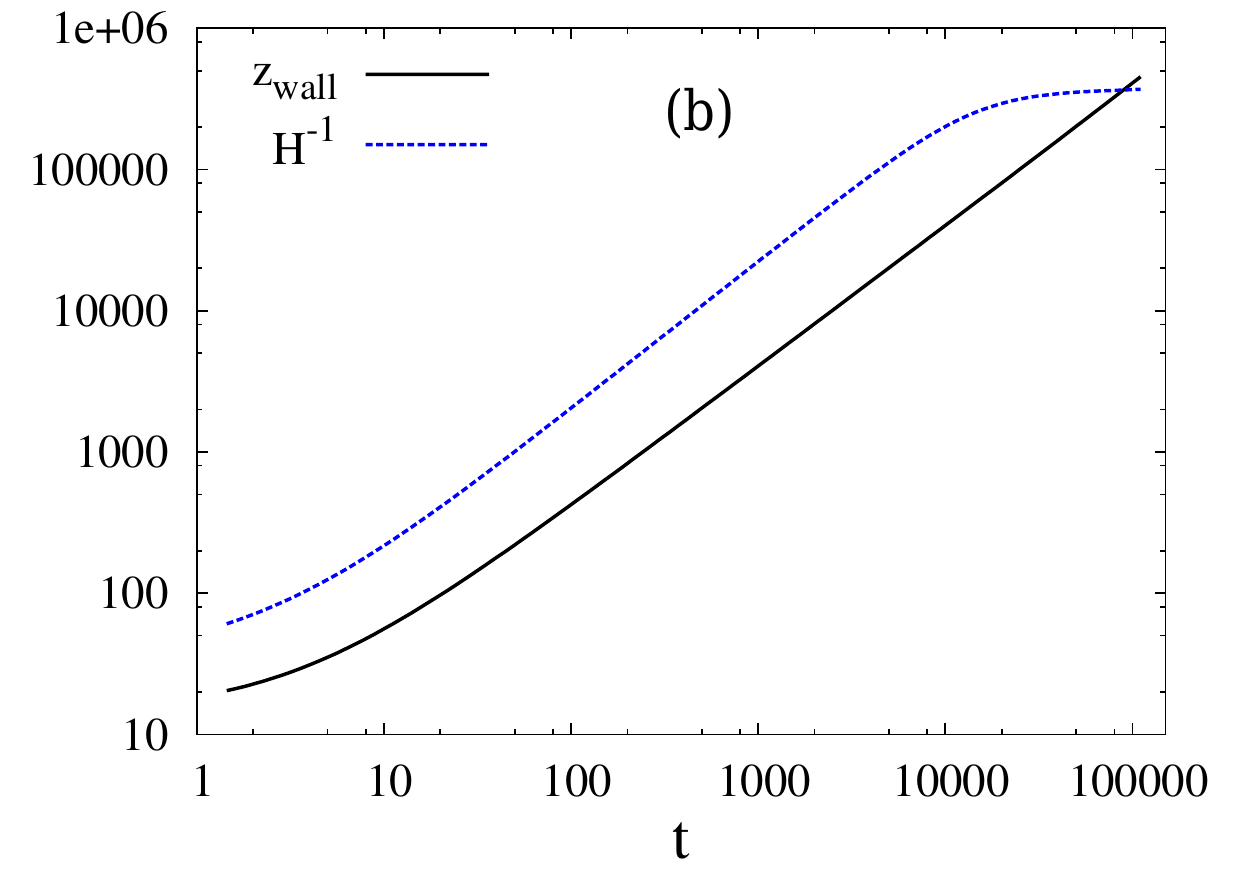}
\caption{ Solid (black) curve shows the location of the domain wall
in physical coordinates, by noting
down the position of the wall where $\phi = 0.25 $vev. Dashed (blue)
curve shows the value of $H^{-1}$. (a) During initial stages, field domain 
expansion is insignificant compared  to the
increase in the value of $H^{-1}$. (b) Towards the end of simulation,
$H^{-1}$ is practically constant while physical size of the domain
(due to expansion of the Universe) keeps increasing. By the end of
simulation here, at $t \simeq 110600$, one can see that the domain exits 
the horizon, marking beginning of proper inflationary stage.}
\label{fig86}
\end{center}
\end{figure}

 Even though the vacuum energy becomes dominant by $t \simeq 15000$, as 
shown in Fig.\ref{fig76}, and $H$ almost constant, if the field domain
remains within the Hubble size $H^{-1}$ then, in principle, dynamics
of field inside the domain can disturb the inflationary phase. To apply the
standard techniques of calculations of density fluctuations etc. we need the
domain to exit the horizon size so that proper $60~ e-fold$ inflation can
occur. We have carried out the simulation until a stage so that 
the stretching of domain by expansion completely overtakes the
increase in $H^{-1}$ (as $H$ becomes almost constant at late stages). 
This is shown in Fig.\ref{fig86}. Solid (black) curve shows the location of 
the domain wall in physical coordinates, 
by noting down the position of the wall where $\phi = 0.25 
$ vev. Dashed (blue) curve shows the value of $H^{-1}$. During initial 
stages, field domain stretching due to universe expansion is 
insignificant compared to the increase in the value of 
$H^{-1}$ as shown in Fig.7a. However, towards the end of simulation,
$H^{-1}$ becomes practically constant while physical size of the domain
(due to expansion of the Universe) keeps increasing. By the end of
simulation here, at $t \simeq 110600$, as shown in Fig.7b, one can see that 
the domain exits the horizon, marking beginning of proper inflationary stage.

\begin{figure}[h]
\begin{center}
\includegraphics[width=0.5\textwidth]{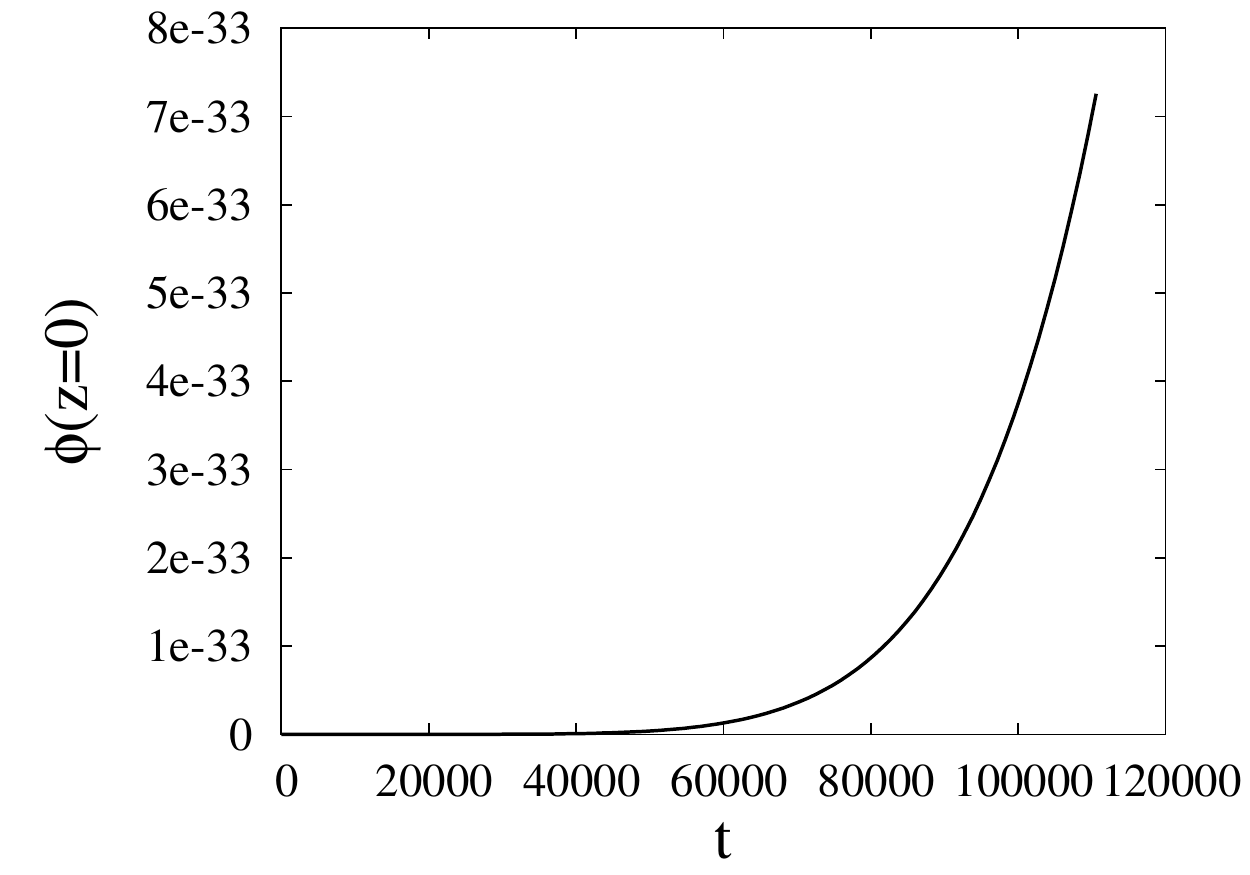}
\caption{The evolution of $\phi(z=0)$.
From this plot, one can make a rough estimate of time after which the field
will become significantly non-zero, ending the period of inflation. A rough 
estimate (assuming linear shape for the final stages in this plot, which may
be a serious overestimate given the shape of the curve, which may very well
be exponential) will 
suggest that after time of order $t \sim 10^{36}$, $\phi(z = 0)$ will become of 
order 1. This period of inflation crucially depends on the size $z_0$ in 
the $\phi$ profile in Eq.\ref{tanh6}. For a larger value of $z_0$ duration of 
inflation will be larger, though at the same time it will become harder to 
justify a very large value of $z_0$.  }
\label{fig96}
\end{center}
\end{figure}

 The duration of inflation will be decided by the roll down of the field
at the center of the domain. With the profile of $\phi$ as given above
for Eq.\ref{tanh6}, $\phi$ is very close to zero at $z = 0$. However, it 
gets pulled down during evolution due to non-zero slope of the potential
as well as due to the non-trivial profile of $\phi$. Fig.\ref{fig96} shows 
the evolution of $\phi(z=0)$. (We show plot on linear scale  here to focus 
on late part of the evolution.) From this, one can estimate the time after 
which the field will become significantly non-zero, causing it to rapidly 
roll down, thereby ending the period of inflation. A rough estimate,
assuming linear shape for the final stages in Fig.8, will suggest
that after time of order $10^{36}$, $\phi(z = 0)$ will become of order 1.
(This may be a serious overestimate given the shape of the plot. It may
appear that the roll down is exponential in  nature. However, the roll
down of the field at center here is governed by complex factors 
depending on the gradient of the field and potential shape. It is thus
not clear that the nature of roll down remains same at later stages.
Thus we do not attempt to fit any specific function to this plot for
determining the duration of inflation.)
This period of inflation crucially depends on the size $z_0$ in the $\phi$
profile in Eq.\ref{tanh6}. For a larger value of $z_0$ duration of 
inflation will be larger, though at the same time it will become harder 
to justify a very large value of $z_0$.

\subsection{3-D field profile}

 As we mentioned above, due to requirements of fixed boundary conditions
for the numerical evolution of field equations, 1-D case has advantage
as one can properly study the evolution of field at the center of
the domain which is crucially important for the total duration of inflation.
Having done that in the previous section, we should now study the 3-D case.
This is particularly important as the nature of propagating front solution
of RD equations crucially depend on the form of the equation.  For the 3-D 
case, for spherically symmetric field profile, the field equation, 
Eq.(15) becomes:

\begin{equation}
\ddot{\phi} + (3H + \eta) \dot{\phi} + \frac{\Lambda^2}{f^2}sin(\phi) - 
{1 \over a^2} (\frac{d^2\phi}{dr^2} + {2 \over r} {d\phi \over dr}) = 0. 
\end{equation}

\begin{figure}[h]
\begin{center}
\includegraphics[width=0.5\textwidth]{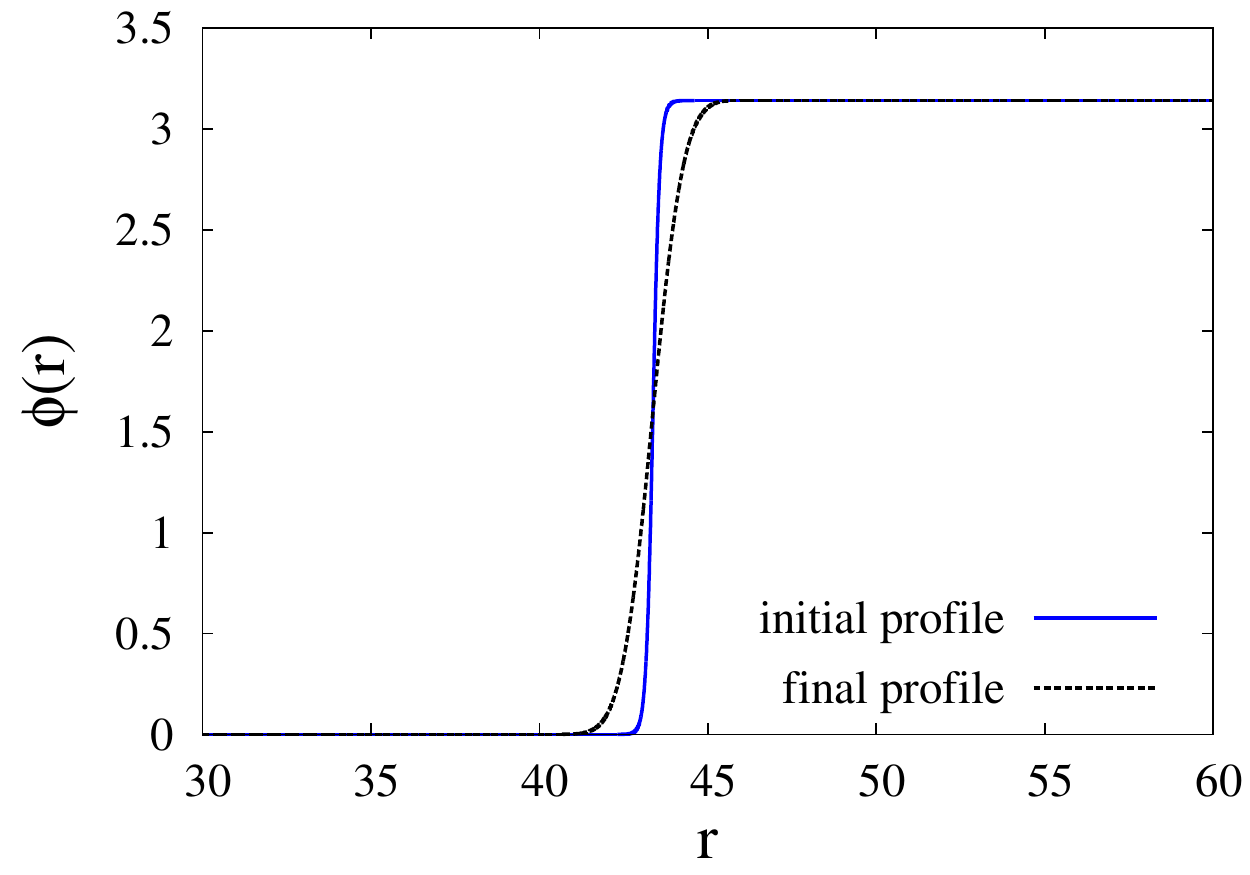}
\caption{The solid (blue) curve shows the initial radial profile of the 
field for the 3-D field profile case. Final plot of $\phi$ (in comoving 
coordinates) at $t =  9348$ is shown by the dashed (black) curve.}
\label{106}
\end{center}
\end{figure}

\begin{figure}[h]
\begin{center}
\includegraphics[width=0.5\textwidth]{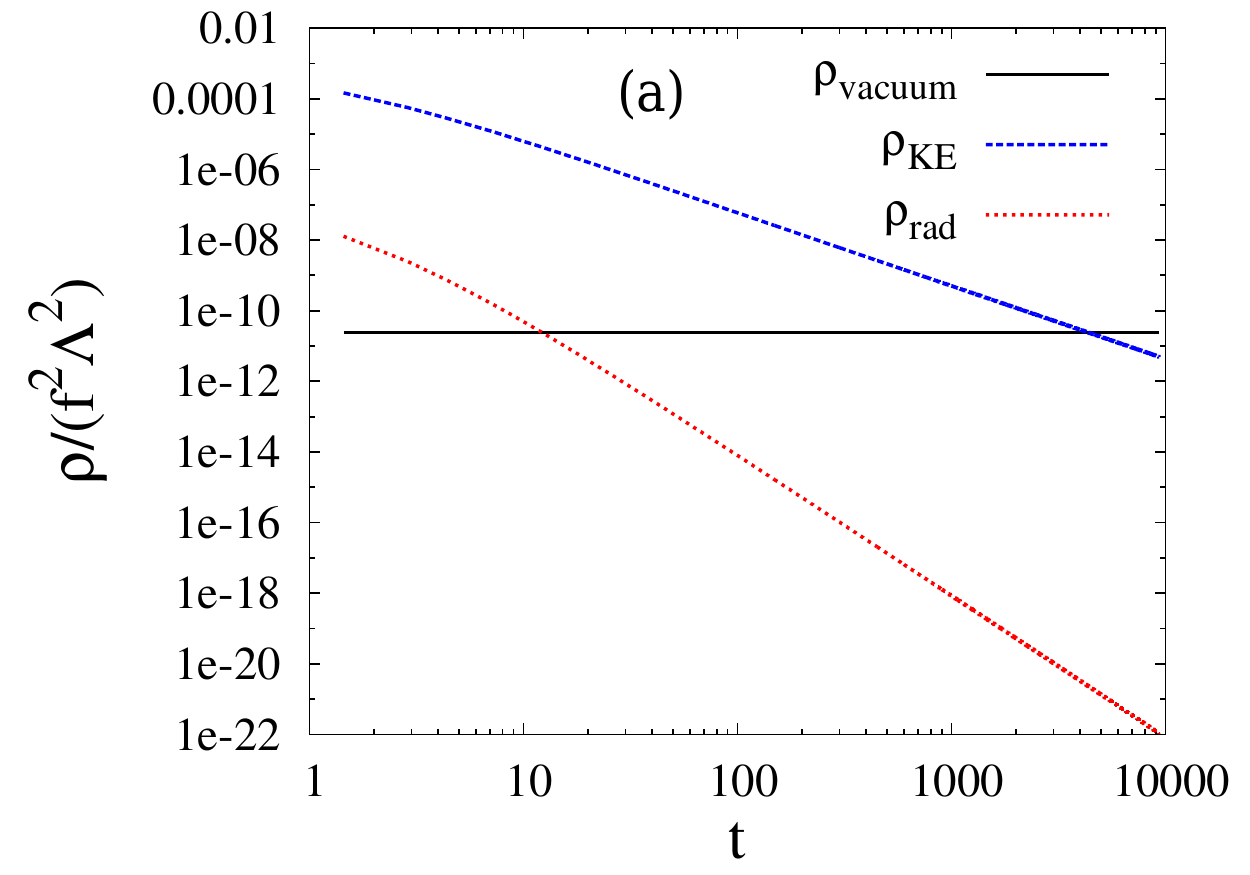}
\includegraphics[width=0.5\textwidth]{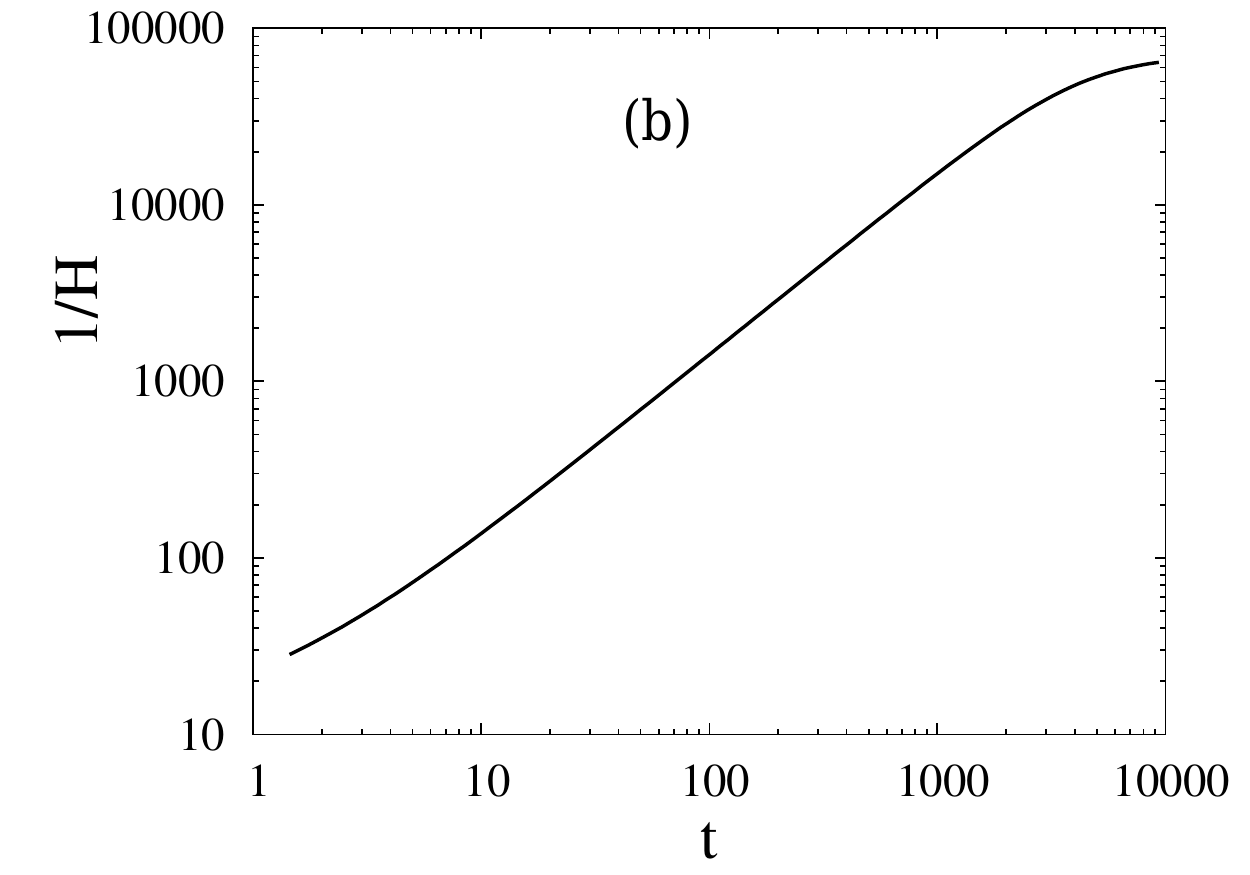}
\caption{(a)Evolution of different components of the energy density
for the 3-D field profile case.
Solid (black), dotted (red), dashed (blue)
curves show the contributions of the field vacuum energy, the field
kinetic energy + gradient energy (denoted by $\rho_{KE}$), and the radiation
energy density respectively. The simulation is run until the vacuum 
energy becomes most dominant component indicating the beginning
of inflationary phase.  (b) Evolution of the Hubble size $H^{-1}$. 
Again, note the turning of the curve from linear
shape  when vacuum energy starts dominating indicating the beginning
of inflationary phase.}
\label{fig116}
\end{center}
\end{figure}

Solutions of this 3-D RD equation can have very different propagating fronts.
For example, in flat space one does not expect any static solution of
3-D RD equation (with standard kinetic terms, due to the well known Derrick's 
theorem). Indeed, we find that the profile of the propagating front is
wider for the 3-D case as compared to the 1-D case (with the same parameter
values). Due to this one needs to have a larger domain size. The radial
field profile at the initial stage (GUT scale) is shown by the solid (black)
curve in Fig.9 (plotting the region near the boundary of the domain). 
Here also we take a $tanh$ profile of the same form
as in 1-D case with $z_0$ giving the size of the region where $\phi$
is close to zero. The diameter of the {\it field domain} in this case is 
taken to be about 87. Again, as for the 1-D case, $H_{GUT}^{-1} \simeq 723$ for 
GUT scale of $10^{15}$ GeV, but incorporating field energy contributions,
the actual value of $H^{-1} \simeq 11$. So, in this case the domain size
is much larger than the actual $H^{-1}$, though still smaller than
$H_{GUT}^{-1}$. We again mention that the domain sizes used here
should be taken as examples where inflation is shown to occur.
It is quite possible that even smaller
domain sizes (here as well as for the 1-D profile case) may still lead to
inflationary phase (especially if larger dissipation terms are chosen). 
The dashed (red) curve in Fig.\ref{106} shows the final profile of $\phi$
at the end of simulation. We have run this simulation only until the 
stage when the vacuum energy starts dominating over the other contributions.
Fig.\ref{fig116} shows plots of various components of energy density, as well
as the plot of the Hubble constant. 
Following the case of 1-D profile, it is clear
that the region will enter inflationary phase. Note, in this case we are
not able to study the roll down of the field at the center of the domain
as field at $r = 0$ is fixed to be $\phi = 0$.

\subsection{Field evolution in the absence of any thermal dissipation}

 We have argued above that it is perfectly reasonable to take thermal
dissipation term in field equations. However, it is important to see whether
in our model, inflationary phase can occur even in the absence of such
a thermal dissipation. For this, we take a 1-D field profile given by
Eq.\ref{tanh6} and evolve it with field equations without any thermal dissipation.
So the field equation is:

\begin{equation}
\ddot{\phi} + 3H\dot{\phi} + \frac{\Lambda^2}{f^2}sin(\phi) - 
{1 \over a^2}{d^2\phi \over dz^2} = 0
\end{equation}

As the dissipation term is much smaller here, field evolution develops
strong oscillations, which is very different from the steadily moving
propagating fronts of RD equation. However, as was discussed in ref.
\cite{our16,our26,skthesis}, 
rough structure of propagating front still remains. In fact,
apart from some oscillations of the field at the front position, remaining
evolution is again by slow motion of the front position. Thus one will
expect that one should again be able to get inflationary phase, though
in this case the vacuum energy needs to dominate over a much larger
contributions from the gradient and kinetic energy contributions
resulting from oscillations. 

\begin{figure}[h]
\begin{center}
\includegraphics[width=0.5\textwidth]{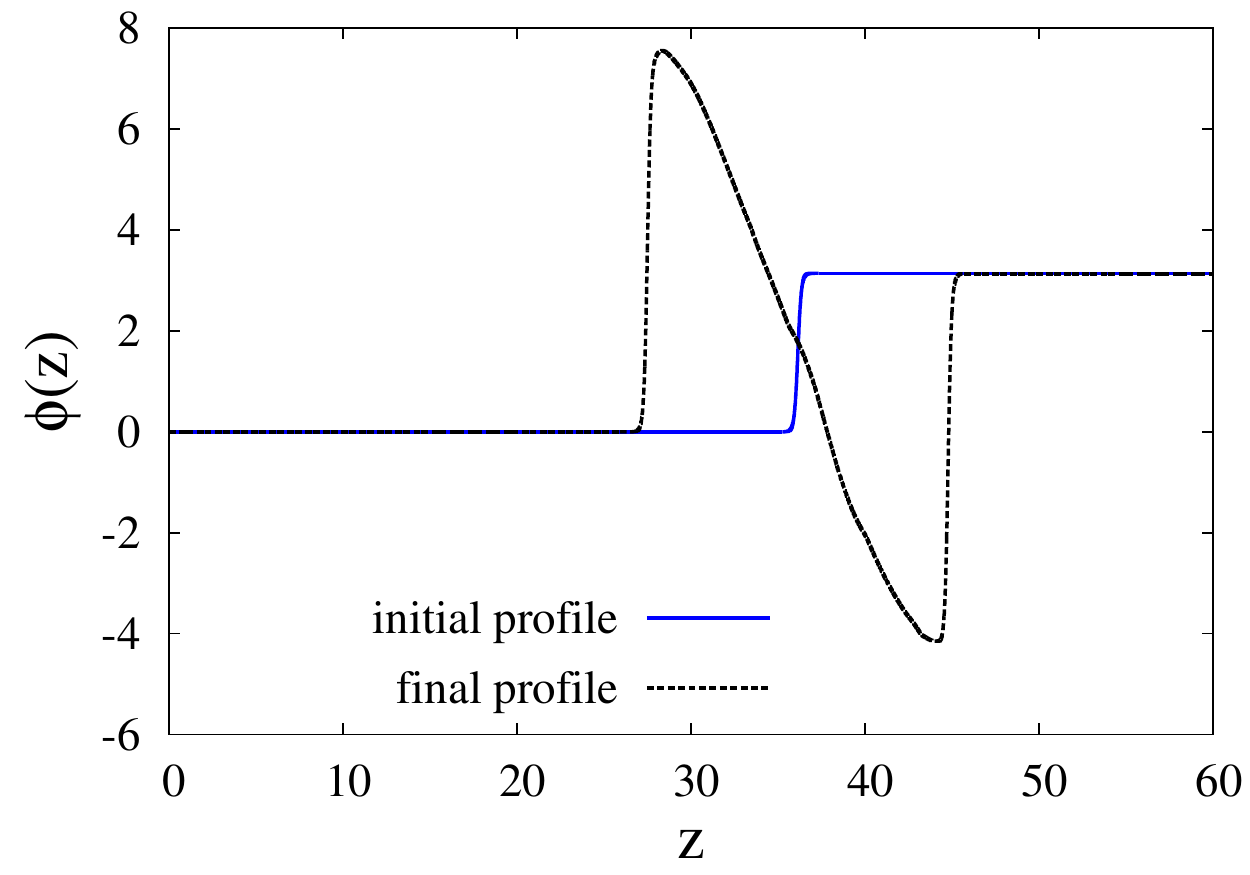}
\caption{Evolution of 1-D field profile for the case without any thermal
dissipation. We only show plot for $z \ge 0$, the profile being 
symmetric about $z = 0$. The solid (blue) curve shows the initial radial 
profile of the field. Final plot of $\phi$ (in comoving coordinates) at 
$t = 14430$ is shown by the dashed (black) curve.}
\label{fig126}
\end{center}
\end{figure}

\begin{figure}[h]
\begin{center}
\includegraphics[width=0.5\textwidth]{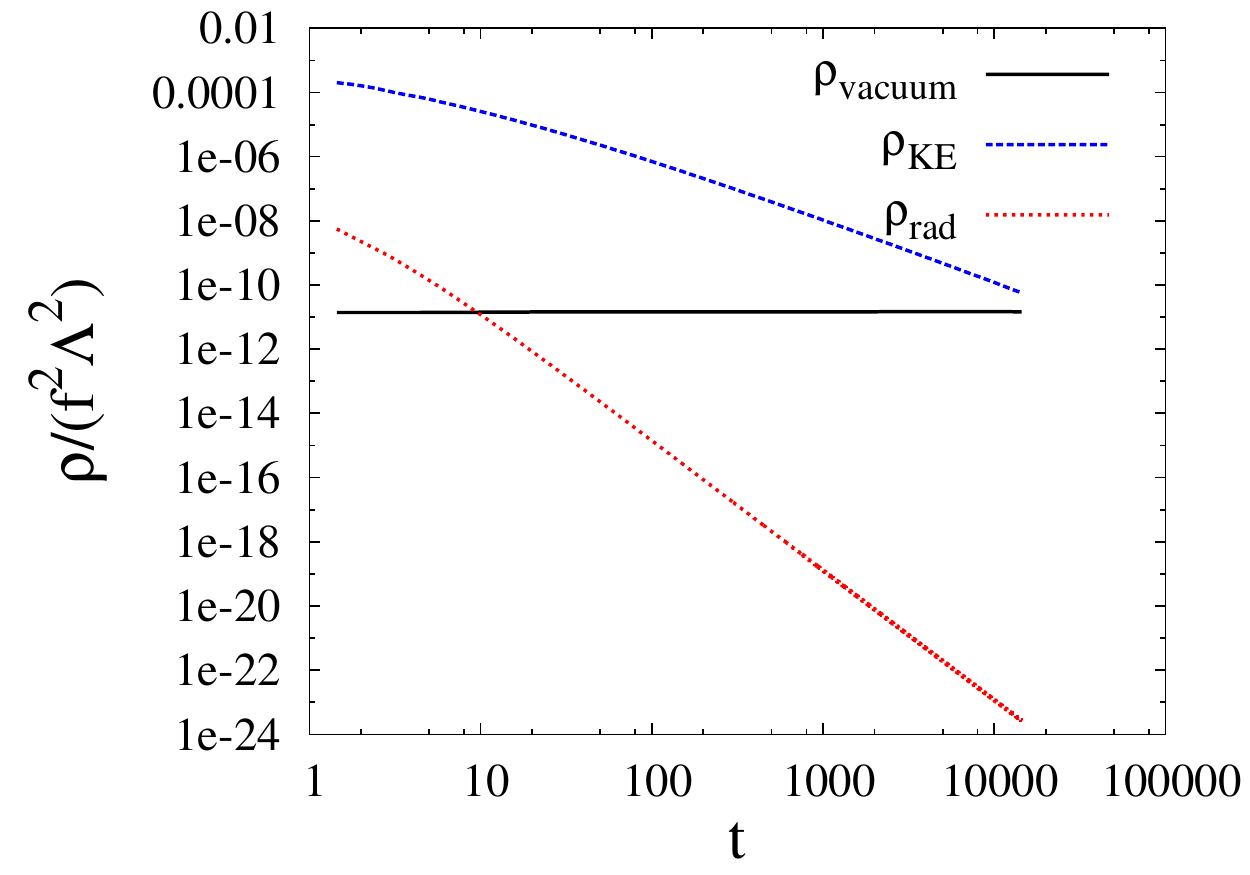}
\caption{Evolution of different components of the energy density
for the case without any thermal dissipation.
Solid (black), dotted (red), dashed (blue)
curves show the contributions of the field vacuum energy, the field
kinetic energy + gradient energy (denoted by $\rho_{KE}$), and the radiation
energy density respectively. The simulation is run until the kinetic
energy + gradient energy becomes close to the vacuum  energy part.}
\label{fig136}
\end{center}
\end{figure}

 Due to smaller dissipation, we need to take larger domain in this case
 (compared to the 1-D case with thermal dissipation).
The diameter of the {\it field domain} in this case is taken to be
about 73, which is 1/10 of $H_{GUT}^{-1} \simeq 723$, but much larger
than actual $H^{-1} \simeq 13$ when field energy contributions
are incorporated. Solid (black) curve in Fig.\ref{fig126} shows this initial
field profile and the dashed (red) curve shows the final profile of $\phi$
at the end of simulation. The final shape of the profile
is almost stable and changes very little subsequently. In this case the
kinetic energy and gradient energy of the field change much more slowly. 
In fact the kinetic energy and gradient energy contributions are almost 
the same here (while in other cases the final kinetic energy was much smaller
than the gradient energy).  We have run this simulation until the 
stage when the kinetic energy + gradient energy becomes close to the 
vacuum energy. It is clear from the plots that the vacuum energy part will
become completely dominant as in Fig.6a, leading to the inflationary phase.
Fig.\ref{fig136} shows plots of various components of energy density, as
well as evolution of $H$.
Following the case of 1-D profile, it is clear
that the region will enter inflationary phase. 

\section{DISCUSSIONS AND CONCLUSIONS}

 We have studied evolution of a highly inhomogeneous field configuration,
forming a sub-horizon domain like structure, as one may
expect to arise after a phase transition, and have shown that in
the presence of thermal dissipation a single such domain can lead
to inflation. This happens due to very special nature of propagating
front solutions of RD equations which slows down field roll down 
sufficiently, allowing domain structure to survive for long time. With this, 
the expansion of the Universe is able to stretch the domain so that the 
vacuum energy starts dominating over other forms of energy densities signaling
beginning of inflation.  The domain we take is still larger than the thermal 
correlation length, but much smaller than $H_{GUT}^{-1}$, and even smaller
than the Hubble parameter incorporating field energy contributions (for
the 1-D profile case, and with thermal dissipation). In the absence of 
thermal dissipation, (as well as for a 3-D field profile) we needed to take 
larger domains, but still one is able to achieve the stage of dominance 
of vacuum energy. Still, for the 3-D case and 1-D case 
without thermal dissipation, as the initial size of the domain is larger 
than the Hubble size, one may say that it looks just like the standard 
treatment where one neglects the field gradients. To certain extent this
would be a correct objection. However we would like 
to note that even in this case, field
gradients play crucial roll in delaying the field roll down at the edges
of the domain. If that did not happen, then the roll down will start
at the edges and may affect the field in inner regions as well, especially
as the domain is not too much larger than the Hubble size. We also emphasize
again (as was mentioned above) that we have not been able to optimize the 
domain sizes for each case due to very long simulations times needed for
each simulation. So, the possibility cannot be ruled out that even for these 
cases required domain size may actually be smaller than the initial Hubble
size.

 One of the most important limitations of our analysis is that we have
not accounted for self gravity of these domains which will obviously
affect the evolution of these domains. We take FRW metric even in the
presence of strongly inhomogeneous energy density. 
However, for the highly inhomogeneous case we are considering, we cannot
say whether the evolution of the whole domain does not change
significantly with a complete treatment of inhomogeneity.  To take care of 
this one needs to have a more rigorous simulation with full Einstein's 
equations as was done in ref.\cite{east26}.  The evolution
of the whole field domain will certainly be very sensitive to the
departure from FRW equations, and one cannot say whether the whole domain
will expand or contract when full account of inhomogeneity is taken.
Still, we expect that the field roll down part may be similar as obtained
from the RD equation front. We do not have strong argument for this, but
if the form of the local field equations remains similar then the only
effect can be to affect the value of the Hubble parameter. (If the
form of field equations also changes significantly then even the
existence of the slowly moving propagating front may be questionable.)
Note that the front thickness is very small compared to the Hubble size
(Fig.5) and the motion of the front from roll down of
the field at that location  is entirely governed by the local field equation.
Though we note that even local field evolution part may be uncertain as the 
value of Hubble constant is needed for the evolution. However, for those 
cases where we have used large thermal dissipation constant $\eta$, the 
value of Hubble constant is not that crucial.

 The requirement of strong dissipation in our model brings to attention the 
warm inflation scenario \cite{warm6}. There is strong dissipation 
present in the warm inflation case
which is responsible for the constant presence of  thermal bath. Study
of RD equation solutions in warm inflation will certainly be very interesting
and may provide new possibilities. At the same time, it also points to another
limitation of our analysis, that is the neglect of fluctuation terms. Certainly,
fluctuations will affect these propagating fronts, as we had also noted in 
ref.\cite{our16,our26,skthesis}. 
We hope to incorporate these effects in a future work.

\acknowledgments

 We are very grateful to Raghavan Rangarajan for very useful and detailed
comments on the manuscript. We also thank Subhendra Mohanty, Nirupam Dutta, 
Oindrila Ganguly, Pranati Rath, and Biswanath Layek for useful
discussions. Some of the results here were presented by AMS
at the International conference  "Saha Theory Workshop: Aspects of Early 
Universe Cosmology", SINP, Kolkata, 16-20 Jan, 2017. We thank the 
participants of this conference, especially Arjun Berera, Koushik Dutta,
and L. Sriramkumar for very useful comments and suggestions. We
thankfully acknowledge Robert Brandenberger for informing us about 
important previous works relating to the issue of initial conditions.

\end{document}